\documentclass[a4paper]{article}
\usepackage{amsfonts}
\usepackage{amssymb}
\usepackage{amsmath}
\usepackage[dvips]{graphicx}
\usepackage{appendix}
\usepackage[a4paper,ignoreall,nomarginpar,twocolumn=false]{geometry}
\usepackage{amstext}
\usepackage[francais]{babel}

\begin{document}

\title{Diffusion d'ondes \'{e}lastiques antiplanes par des r\'{e}seaux p\'{e}%
riodiques bidimensionnels de fissures\bigskip \\
Scattering of antiplane elastic waves by two-dimensional periodic arrays of
cracks\medskip }
\author{Mihai Caleap \\
Department of Mechanical Engineering, University of Bristol, \\
Queen's Building, University Walk, Bristol BS8 1TR, UK}
\maketitle

\textbf{R\'{e}sum\'{e}}. Dans le contexte de la propagation d'ondes \'{e}%
lastiques dans des milieux endommag\'{e}s, une approche analytique pour la
diffusion d'ondes antiplanes par des r\'{e}seaux p\'{e}riodiques
bidimensionnels de fissures est d\'{e}velopp\'{e}e. Avant d'envisager l'\'{e}%
tude de r\'{e}seaux de fissures, la diffusion simple d'une onde \'{e}%
lastique antiplane par une fissure plate est pr\'{e}alablement \'{e}tudi\'{e}%
e. Ensuite, en employant la formule de repr\'{e}sentation d'une fissure
plate et la condition de p\'{e}riodicit\'{e} d'une rang\'{e}e de fissures
situ\'{e}es sur la m\^{e}me droite antiplane, une \'{e}quation int\'{e}grale
de fronti\`{e}res est alors obtenue pour le pour le saut de d\'{e}placement 
\`{a} travers une fissure dite de r\'{e}f\'{e}rence. Des r\'{e}sultats num%
\'{e}riques pour les coefficients de r\'{e}flexion et de transmission sont pr%
\'{e}sent\'{e}s en fonction de l'espacement entre fissures, de la fr\'{e}%
quence d'excitation et de l'angle d'incidence. Finalement, l'\'{e}tude de la
propagation d'une onde antiplane au travers d'un empilement p\'{e}riodique
d'un nombre fini de r\'{e}seaux lin\'{e}aires (i.e. un r\'{e}seau dit
bidimensionnel p\'{e}riodique) est ax\'{e}e sur l'analyse des bandes
spectrales interdites et permises. Les effets dus \`{a} une perturbation de
la p\'{e}riodicit\'{e} ont \'{e}galement \'{e}t\'{e} analys\'{e}s.\bigskip

\textbf{Abstract.} In the context of elastic wave propagation in damaged
solids, an analytical approach for scattering of antiplane waves by
two-dimensional periodic arrays of cracks is developed. Before considering
the study of arrays of cracks, the scattering of an antiplane wave by a flat
crack is first studied. Then, using the representation formula for the
scattered displacement by a flat and by considering the periodicity
condition of the crack-spacing, a boundary integral equation is obtained for
the crack face displacement of the reference crack. Numerical results for
the reflection and transmission coefficients are presented as functions of
the crack-spacing, the frequency of excitation, and the angle of incidence.
Finally, the propagation of antiplane waves by two-dimensional periodic
arrays of cracks is studied. Despite the use of a finite number of linear
arrays, one recognizes the effects of band-pass filtering or band rejection
characteristics of the transmission spectra of a periodic medium. Effects
due to a disorder in the periodicity are also analysed.\bigskip

\textit{Mots-cl\'{e}s :}\textbf{\ }endommagement ; r\'{e}seau de fissures~;
propagation d'ondes \'{e}lastiques\bigskip

\textit{Keywords:}\textbf{\ }damage; array of cracks; elastic wave
propagation\newpage

\section*{Abridged English version}

Localized damage in form of distributed coplanar micro-cracks often occurs
at the bounding surfaces between two materials of the same or different
mechanical properties. Such micro-cracks may be induced by manufacturing,
materials processing, or under service conditions. The nucleation, growth
and coalescence of that localized damage may gave rise to substantial
stiffness and strength reductions of the materials. Thus, the detection and
characterization of the damage state in an engineering structure is of
considerable interests for improving the manufacturing and the material
processing, estimating the life-time, and avoiding catastrophic failures of
a damaged structure. This research has also considerable interest, on the
one hand to ultrasonic quantitative non-destructive evaluation for detecting
and characterizing the damaged state of composites, and on the other hand to
linear elastic fracture mechanics to assess the initiation and growth of a
developed macro-crack under dynamic loading conditions. With this
motivation, the present analysis is concerned with elastic antiplane (shear
horizontal) wave propagation in isotropic materials containing micro-cracks
arranged periodically in two space-directions. Although an antiplane wave
has limited applications in ultrasonic non-destructive evaluation, some
useful information can be extracted from this simple model.

\subsection*{\textit{Summary}}

A method similar to that proposed by Zhang and Gross \cite{ZG} is used to
formulate the problem of scattering by a periodic array of collinear shear
cracks. The geometry of the problem is presented in Fig. 1. By using a
representation formula for the scattered displacement and by considering the
periodicity condition of the crack-spacing, an integral equation is obtained
for the crack face displacement of a reference crack, which is solved
numerically by using a Galerkin method. In general, the scattered wave field
consists of an infinite number of wave modes, each with its own cut-off
frequency. Under the cut-off frequency the wave modes are standing wave
modes which decay exponentially with distance to the array. At frequencies
higher than the cut-off frequency the wave modes will be propagating. At the
cut-off frequency a mode conversion from a standing into a propagating wave
mode occurs. Sufficiently far from the array, and at sufficiently low
frequencies, only the zeroth (lowest) order mode is dominant, and that mode
represents homogeneous plane waves corresponding to reflected and
transmitted waves by the cracks. Numerical results for the reflection
(transmission) coefficient are presented, as function of the crack-spacing,
the wave frequency, and the angle of incidence.

The wave propagation in two-dimensional periodic arrays of flat cracks is
studied next. The global array of Fig. 5 is composed of an arbitrary number $%
S$ of infinite periodic arrays of collinear cracks, regularly spaced by a
distance $D$ along the axis $y_{3}$. Its thickness is $h=(S-1)D$. We assume
that all linear arrays are identical, with $d$ their common period (along
the axis $y_{1}$) and $2a$ the crack width. The reflection and transmission
coefficients of the linear array $s$ ($s=1,2,...,S$) are known from the
previous investigation. From these coefficients, the global reflection and
transmission coefficients of the two-dimensional array are calculated
without difficulty by using an iterative algorithm. Other closely related
problems, some more complicated as they involve inplane ((longitudinal or
shear vertical)) waves, can also be found in the literature; see, e.g. \cite%
{Sca1} and \cite{Li}. The formalism outlined in the text is valid only at
low frequencies and only the lowest order modes can propagate in the array.
The other modes are evanescent and decrease exponentially with the
propagation distance. They can therefore be neglected if the distance $D$ is
large enough, such that an evanescent mode can not disturb significantly a
neighbor linear array. Despite the use of a finite number of linear arrays,
one recognizes the effects of band-pass filtering or band rejection
characteristics of the transmission spectra of a periodic medium. The band
rejections occur in the vicinity of frequencies for which there is
coincidence between the spacing of linear arrays and multiples of the
half-wavelength.

Finally, for an infinite periodic two-dimensional array of cracks, the
effective wave number of antiplane waves propagating in the homogeneous
medium equivalent to two-dimensional array is calculated by using the
Floquet theory.

The theory developed here can be further generalized to a two-dimensional
array of cracks with any type of periodicity, such as a variation of the
crack-sizes or a random distribution of linear arrays.

\section{ Introduction}

Un endommagement localis\'{e} sous forme d'une distribution de microfissures
coplanaires se produit souvent au niveau des surfaces de liaison entre deux
mat\'{e}riaux aux propri\'{e}t\'{e}s m\'{e}caniques identiques ou diff\'{e}%
rentes. De telles microfissures peuvent \^{e}tre induites durant la
fabrication, par les mat\'{e}riaux traitants, ou bien les conditions
d'utilisation. La germination, la croissance et la coalescence des dommages
localis\'{e}s peuvent provoquer des r\'{e}ductions substantielles de rigidit%
\'{e} et de r\'{e}sistance des mat\'{e}riaux. Ainsi, la d\'{e}tection et la
caract\'{e}risation de l'\'{e}tat d'endommagement dans une structure d'ing%
\'{e}nierie est particuli\`{e}rement importante pour am\'{e}liorer le
processus de fabrication, les mat\'{e}riaux traitants, et \'{e}viter les
ruptures catastrophiques. Cette recherche a aussi un int\'{e}r\^{e}t consid%
\'{e}rable, d'une part pour l'\'{e}valuation ultrasonore non-destructive,
dans le but de d\'{e}tecter et caract\'{e}riser l'\'{e}tat d'endommagement
de composites, et d'autre part, pour la m\'{e}canique \'{e}lastique lin\'{e}%
aire de la rupture qui permet d'\'{e}valuer le d\'{e}clenchement et la
croissance d'une \guillemotleft ~macro-fissure~\guillemotright\ pouvant se d%
\'{e}velopper dans des conditions de charges dynamiques. Bien que beaucoup
de facteurs puissent surgir \textit{in situ}, les composites
unidirectionnels renforc\'{e}s ainsi que les lamin\'{e}es \`{a} plis crois%
\'{e}s peuvent \^{e}tre commod\'{e}ment approxim\'{e}s comme des mat\'{e}%
riaux \'{e}lastiques \`{a} sym\'{e}trie isotrope ou orthotrope. Une telle
approximation est possible lorsque la longueur d'onde de l'onde incidente
est beaucoup plus grande que les longueurs caract\'{e}ristiques des mat\'{e}%
riaux composites (espacement, diam\`{e}tre des fibres dans un composite
renforc\'{e}, \'{e}paisseur de diff\'{e}rentes couches d'un stratifi\'{e}).

Ainsi, l'analyse de la section 2 concerne la propagation d'une onde
antiplane dans des mat\'{e}riaux isotropes contenant des microfissures dispos%
\'{e}es p\'{e}riodiquement dans une direction spatiale, et formant un r\'{e}%
seau lin\'{e}aire infini. La section 3 a pour objectif l'\'{e}tude de la
propagation d'ondes antiplanes dans des r\'{e}seaux p\'{e}riodiques \`{a}
deux dimensions spatiales. La m\'{e}thode de calcul des coefficients de r%
\'{e}flexion et de transmission du r\'{e}seau bidimensionnel, ainsi que
l'obtention des courbes de dispersion et d'att\'{e}nuation correspondantes,
reposent sur la d\'{e}composition des r\'{e}seaux bidimensionnels en un
nombre fini de r\'{e}seaux lin\'{e}aires dont leurs propres coefficients de r%
\'{e}flexion et de transmission sont d\'{e}sormais connus.

\section{ Arrangement p\'{e}riodique de fissures colin\'{e}aires}

Dans cette section, un mod\`{e}le est pr\'{e}sent\'{e} afin de calculer les
coefficients de r\'{e}flexion et de transmission par un arrangement p\'{e}%
riodique de fissures colin\'{e}aires dans un solide \'{e}lastique isotrope,
pour le cas d'une onde antiplane. M\^{e}me si les ondes antiplanes ont des
applications limit\'{e}es dans l'\'{e}valuation ultrasonore non-destructive,
des informations utiles peuvent \^{e}tre extraites \`{a} partir de ce mod%
\`{e}le simple. Une m\'{e}thode similaire \`{a} celle propos\'{e}e par Zhang
et Gross \cite{ZG} est employ\'{e}e pour formuler le probl\`{e}me de
diffusion par un r\'{e}seau de fissures lin\'{e}aire et infini. Les d\'{e}%
placements diffus\'{e}s sont exprim\'{e}s sous forme d'int\'{e}grales de
Fourier contenant les d\'{e}placements au niveau des l\`{e}vres des
fissures. En employant la formule de repr\'{e}sentation d'une fissure plate
et la condition de p\'{e}riodicit\'{e} d'une rang\'{e}e de fissures situ\'{e}%
es sur la m\^{e}me droite antiplane, une \'{e}quation int\'{e}grale de fronti%
\`{e}res est alors obtenue pour le d\'{e}placement des l\`{e}vres de la
fissure de r\'{e}f\'{e}rence. G\'{e}n\'{e}ralement, le champ de d\'{e}%
placement diffus\'{e} se compose d'un nombre infini de modes d'onde, chacun
avec sa propre fr\'{e}quence de coupure. Au-dessous de leur fr\'{e}quence de
coupure, les modes sont dits ondes stationnaires et d\'{e}croissent
exponentiellement au cours de la propagation. Au-d\'{e}l\`{a} de la fr\'{e}%
quence de coupure, les modes deviennent des ondes homog\`{e}nes. \`{A} la fr%
\'{e}quence de coupure, une conversion du mode stationnaire en un mode homog%
\`{e}ne se produit. Lorsque situ\'{e} suffisamment loin de l'interface, et 
\`{a} des fr\'{e}quences suffisamment basses, le mode d'ordre z\'{e}ro est
dominant, repr\'{e}sentant les ondes planes homog\`{e}nes r\'{e}fl\'{e}chies
et transmises par les fissures. Des r\'{e}sultats num\'{e}riques pour les
coefficients de r\'{e}flexion et de transmission sont pr\'{e}sent\'{e}s en
fonction de l'espacement entre fissures, de la fr\'{e}quence d'excitation et
de l'angle d'incidence.

\subsection{ Champ diffus\'{e} par une fissure}

Consid\'{e}rons un solide \'{e}lastique isotrope, homog\`{e}ne, infini,
contenant une fissure de longueur $2a$. La fissure est centr\'{e}e \`{a}
l'origine du syst\`{e}me $\left( y_{1},y_{3}\right) $ et est infiniment
longue dans la direction perpendiculaire $y_{2}$. Nous soumettons cette
fissure \`{a} un champ acoustique antiplan, d\'{e}fini par un champ de d\'{e}%
placement dans la direction $y_{2}$, de la forme~:

\begin{equation}
u_{2}^{\mathrm{inc}}\left( y_{1},y_{3}\right) =u_{0}\mathrm{e}^{\mathrm{i}%
k_{T}\left( y_{1}\sin \theta _{0}+y_{3}\cos \theta _{0}\right) },  \label{1}
\end{equation}%
o\`{u} $u_{0}$ et $\theta _{0}$ repr\'{e}sentent respectivement l'amplitude
de l'onde et l'angle d'incidence. D'apr\`{e}s la loi de Hooke, la contrainte 
$\sigma _{23}^{\mathrm{inc}}$ associ\'{e}e est donn\'{e}e, dans le plan de
la fissure ($y_{3}=0$), par

\begin{equation}
\sigma _{23}^{\mathrm{inc}}\left( y_{1},0\right) =\mathrm{i}u_{0}k_{T}\mu
_{0}\cos \theta _{0}\mathrm{e}^{\mathrm{i}k_{T}y_{1}\sin \theta _{0}},
\label{2}
\end{equation}%
o\`{u} $k_{T}=\omega /c_{T}$ est le nombre d'onde, $c_{T}$ est la c\'{e}l%
\'{e}rit\'{e} transverse et $\mu _{0}$ est le module de rigidit\'{e} du
solide. Puisque il y a une invariance en la translation dans la direction $%
y_{2}$, le probl\`{e}me de diffusion r\'{e}sultant devient naturellement
bidimensionnel.

Quand la fissure interf\`{e}re avec l'onde incidente, des ondes diffus\'{e}%
es sont g\'{e}n\'{e}r\'{e}es. Ces ondes produisent un d\'{e}placement diffus%
\'{e} antiplan\ $u_{2}^{\mathrm{dif}}$, solution de l'\'{e}quation de
Helmholtz scalaire

\begin{equation}
\frac{\partial ^{2}}{\partial y_{1}^{2}}u_{2}^{\mathrm{dif}}\left(
y_{1},y_{3}\right) +\frac{\partial ^{2}}{\partial y_{3}^{2}}u_{2}^{\mathrm{%
dif}}\left( y_{1},y_{3}\right) +k_{T}^{2}u_{2}^{\mathrm{dif}}\left(
y_{1},y_{3}\right) =0.  \label{3}
\end{equation}%
Notons que le d\'{e}placement diffus\'{e} est antisym\'{e}trique par rapport
au plan $\left( y_{1},y_{2}\right) $. Nous pouvons donc limiter l'\'{e}tude
au demi-espace $y_{3}>0$. En supposant que les l\`{e}vres de la fissure sont
libres de contraintes, la condition aux limites du probl\`{e}me se traduit
par~:

\begin{equation}
\sigma _{23}^{\mathrm{dif}}\left( y_{1},0^{+}\right) =-\mathrm{i}%
u_{0}k_{T}\mu _{0}\cos \theta _{0}\mathrm{e}^{\mathrm{i}k_{T}y_{1}\sin
\theta _{0}},\text{ \ }(\left\vert y_{1}\right\vert <a).  \label{4}
\end{equation}%
En dehors de la fissure, le champ de d\'{e}placement vaut~:

\begin{equation}
u_{2}^{\mathrm{dif}}(y_{1},0^{+})=0,\text{ \ }(\left\vert y_{1}\right\vert
>a).  \label{5}
\end{equation}

\subsubsection{ Repr\'{e}sentation int\'{e}grale}

D\'{e}finissons la transform\'{e}e de Fourier spatiale (par rapport \`{a} la
variable $y_{1}$) directe et inverse de $u_{2}^{\mathrm{dif}}$~:

\begin{subequations}
\begin{equation}
\tilde{u}_{2}^{\mathrm{dif}}\left( \xi ,y_{3}\right) =\frac{1}{\sqrt{2\pi }}%
\int_{-\infty }^{\infty }u_{2}^{\mathrm{dif}}\left( \vec{y}\right) \mathrm{e}%
^{\mathrm{i}\xi y_{1}}\mathrm{d}y_{1},  \label{TF1}
\end{equation}

\begin{equation}
u_{2}^{\mathrm{dif}}\left( \vec{y}\right) =\frac{1}{\sqrt{2\pi }}%
\int_{-\infty }^{\infty }\tilde{u}_{2}^{\mathrm{dif}}\left( \xi
,y_{3}\right) \mathrm{e}^{-\mathrm{i}\xi y_{1}}\mathrm{d}\xi .  \label{TF2}
\end{equation}%
Apr\`{e}s avoir appliqu\'{e} la transform\'{e}e \eqref{TF1} \`{a} l'\'{E}q. %
\eqref{3} et d\'{e}fini la fonction $\beta $ telle que~:

\end{subequations}
\begin{equation}
\beta \left( \xi \right) =\left\{ 
\begin{array}{ll}
{\left( \xi ^{2}-k_{T}^{2}\right) ^{1/2},} & {\xi ^{2}\geq k_{T}^{2},} \\ 
{-\mathrm{i}\left( k_{T}^{2}-\xi ^{2}\right) ^{1/2},} & {\xi ^{2}<k_{T}^{2},}%
\end{array}%
\right.  \label{B}
\end{equation}%
l'\'{e}quation de Helmholtz \eqref{3} devient une \'{e}quation diff\'{e}%
rentielle ordinaire de variable $y_{3}$~:

\begin{equation}
\frac{\partial ^{2}\tilde{u}_{2}^{\mathrm{dif}}}{\partial y_{3}^{2}}\left(
\xi ,y_{3}\right) -\beta ^{2}\left( \xi \right) \tilde{u}_{2}^{\mathrm{dif}%
}\left( \xi ,y_{3}\right) =0.  \label{17)}
\end{equation}%
Une solution born\'{e}e dans la r\'{e}gion $y_{3}>0$ et satisfaisant la
condition de Sommerfeld \cite{Somer}, s'\'{e}crit~:

\begin{equation}
\tilde{u}_{2}^{\mathrm{dif}}\left( \xi ,y_{3}\right) =C\left( \xi \right) 
\mathrm{e}^{-\beta y_{3}},  \label{ZEqnNum276831}
\end{equation}%
o\`{u} la fonction $C$ est donn\'{e}e par l'\'{E}q. \eqref{TF1} sous la
forme~:

\begin{equation}
C\left( \xi \right) =\frac{1}{\sqrt{2\pi }}\int_{-a}^{a}u_{2}^{\mathrm{dif}%
}\left( y_{1},0^{+}\right) \mathrm{e}^{\mathrm{i}\xi y_{1}}\mathrm{d}y_{1}.
\label{ZEqnNum755537}
\end{equation}%
Nous pouvons maintenant substituer l'\'{E}q. \eqref{ZEqnNum755537} dans l'%
\'{E}q. \eqref{ZEqnNum276831} et appliquer la transform\'{e}e inverse %
\eqref{TF2}. La solution du probl\`{e}me de diffusion \eqref{4}-\eqref{5}
peut s'\'{e}crire sous la forme~:

\begin{equation}
u_{2}^{\mathrm{dif}}\left( y_{1},y_{3}\right) =\frac{1}{2\pi }\mathrm{sgn}%
\left( y_{3}\right) \int_{-\infty }^{\infty }\int_{-a}^{a}u_{2}^{\mathrm{dif}%
}\left( v,0^{+}\right) \mathrm{e}^{\mathrm{i}\xi \left( v-y_{1}\right) }%
\mathrm{e}^{-\beta \left( \xi \right) \left\vert y_{3}\right\vert }\mathrm{d}%
v\mathrm{d}\xi .  \label{6}
\end{equation}%
L'\'{e}quation \eqref{6} montre qu'il est possible d'\'{e}valuer $u_{2}^{%
\mathrm{dif}}$ quel que soit $\left( y_{1},y_{3}\right) \in 
\mathbb{R}
^{2}$, si $u_{2}^{\mathrm{dif}}\left( v,0^{+}\right) $ est connu pour $%
\left\vert v\right\vert <a$. En utilisant l'\'{E}q. \eqref{6} dans la
condition aux limites \eqref{4}, nous obtenons l'\'{e}quation integrale
suivante pour $u_{2}^{\mathrm{dif}}\left( v,0^{+}\right) $ :

\begin{equation}
\int_{-a}^{a}u_{2}^{\mathrm{dif}}\left( v,0^{+}\right) \int_{-\infty
}^{\infty }\beta \left( \xi \right) \mathrm{e}^{\mathrm{i}\xi \left(
v-y_{1}\right) }\mathrm{d}\xi \mathrm{d}v=k_{T}\cos \theta _{0}\mathrm{e}^{%
\mathrm{i}k_{T}y_{1}\sin \theta _{0}},\text{ \ }(\left\vert y_{1}\right\vert
<a).  \label{int}
\end{equation}%
Angel \cite{Angel} a montr\'{e} que l'int\'{e}grale double dans le membre de
gauche de l'\'{E}q. \eqref{int} se r\'{e}duit \`{a} une int\'{e}grale singuli%
\`{e}re de type Cauchy. L'\'{e}quation integrale \eqref{int} ne peut pas 
\^{e}tre r\'{e}solue analytiquement, mais il est possible de la r\'{e}soudre
num\'{e}riquement en utilisant, par exemple, une formule d'int\'{e}gration
de type Gauss-Tchebychev (voir \cite{Clp}, chapitre 2, pour plus de d\'{e}%
tails).

\subsubsection{ Champ lointain}

Soit $\left( r,\theta \right) $ les coordonn\'{e}es polaires d\'{e}finies
par $r^{2}=y_{1}^{2}+y_{3}^{2}$ et $\cos \theta =y_{3}/r.$ En champ
lointain, lorsque $r\gg a$, le deplacement diffus\'{e} \eqref{6} peut s'\'{e}%
crire comme le produit d'une fonction radiale, qui d\'{e}pend de $k_{T}r$,
et d'une fonction de l'angle d'observation $\theta $, qui d\'{e}pend \textit{%
a priori} de l'angle d'incidence $\theta _{0}$\ et de la fr\'{e}quence. Nous
avons \cite{CAA} :

\begin{equation}
u_{2}^{\mathrm{dif}}\left( r,\theta ,\theta _{0}\right) =u_{0}\sqrt{\frac{2%
\mathrm{i}\pi }{k_{T}r}}\mathrm{e}^{\mathrm{i}k_{T}r}f\left( \theta ,\theta
_{0}\right) ,\text{ \ }\left( r\gg a,\text{ \ }\theta \in \lbrack 0,2\pi
)\right) ,
\end{equation}%
o\`{u} l'\textit{amplitude de diffusion en champ lointain }$f$ est d\'{e}%
finie par :

\begin{equation}
f\left( \theta ,\theta _{0}\right) =k_{T}\cot \theta \int_{-a}^{a}u_{2}^{%
\mathrm{dif}}\left( v,0^{+}\right) \mathrm{e}^{-\mathrm{i}k_{T}v\sin \theta }%
\mathrm{d}v.  \label{Ftheta}
\end{equation}

\subsection{ Champ diffus\'{e} par un arrangement p\'{e}riodique de fissures
colin\'{e}aires}

L'objectif de cette section est d'\'{e}crire le champ diffus\'{e} par un r%
\'{e}seau lin\'{e}aire infini sous la forme d'ondes planes r\'{e}fl\'{e}%
chies et transmises. La g\'{e}om\'{e}trie du r\'{e}seau et les notations
utilis\'{e}es sont pr\'{e}sent\'{e}es sur la Fig. 1.

\subsubsection{ \'{E}nonc\'{e} du probl\`{e}me}

Le r\'{e}seau est compos\'{e} d'une disposition p\'{e}riodique de fissures
antiplanes colin\'{e}aires. Soit $2a$ la longueur de chaque fissure et $d$
la distance entre les centres de deux fissures voisines (pas de r\'{e}seau).
Soit $\left( \zeta _{1},\zeta _{3}\right) $ le rep\`{e}re \guillemotleft %
~local~\guillemotright\ li\'{e} \`{a} une fissure. Le rep\`{e}re 
\guillemotleft ~global~\guillemotright\ $\left( y_{1},y_{3}\right) $ est reli%
\'{e} au rep\`{e}re local $\left( \zeta _{1},\zeta _{3}\right) $ par~:

\begin{equation}
\left( 
\begin{array}{c}
{y_{1}} \\ 
{y_{3}}%
\end{array}%
\right) =\left( 
\begin{array}{c}
{\zeta _{1}} \\ 
{\zeta _{3}}%
\end{array}%
\right) +\left( 
\begin{array}{c}
{jd} \\ 
{0}%
\end{array}%
\right) ,\text{ \ }\left\vert \zeta _{1}\right\vert \leq a.  \label{8}
\end{equation}

D'apr\`{e}s l'\'{E}q. \eqref{6}, le champ de d\'{e}placement diffus\'{e} au
point d'observation $\left( y_{1},y_{3}\right) $ est donc donn\'{e} par~:

\begin{equation}
u_{2}^{\mathrm{dif}}\left( y_{1},y_{3}\right) =\frac{1}{2\pi }\mathrm{sgn}%
\left( y_{3}\right) \sum_{j=-\infty }^{\infty }\int_{-\infty }^{\infty
}\int_{-a}^{a}u_{2}^{j}\left( \zeta _{1},0^{+}\right) \mathrm{e}^{{\mathrm{i}%
}\xi \left( \zeta _{1}+jd-y_{1}\right) }\mathrm{e}^{-\beta \left( \xi
\right) \left\vert y_{3}\right\vert }\mathrm{d}\zeta _{1}\mathrm{d}\xi ,
\label{9}
\end{equation}%
o\`{u} $u_{2}^{j}\left( \zeta _{1},0^{+}\right) $ est le d\'{e}placement
diffus\'{e} sur la fissure $j$.

\subsubsection{ Hypoth\`{e}se de p\'{e}riodicit\'{e}}

D'une fissure \`{a} une autre le d\'{e}placement incident est le m\^{e}me
mais d\'{e}phas\'{e} de la quantit\'{e} $k_{T}d\sin \theta _{0}$. Ce terme
vaut $1$ quand $\theta _{0}=0$, c'est-\`{a}-dire en incidence normale, quand
toutes les fissures sont excit\'{e}es en m\^{e}me temps. En substituant %
\eqref{8} dans \eqref{2}, le champ de contrainte incident sur les l\`{e}vres
des fissures peut s'\'{e}crire comme~:

\begin{equation}
\sigma _{23}^{\mathrm{inc}}\left( y_{1},0\right) =\overset{\frown }{\sigma }%
_{23}^{\mathrm{inc}}\left( \zeta _{1}\right) \mathrm{e}^{\mathrm{i}%
k_{T}jd\sin \theta _{0}},  \label{10}
\end{equation}%
o\`{u} $\overset{\frown }{\sigma }_{23}^{\mathrm{inc}}\left( \zeta
_{1}\right) $ est le \textit{champ de contrainte incident sur la fissure de r%
\'{e}f\'{e}rence} ($\left\vert \zeta _{1}\right\vert \leq a$, $\zeta _{3}=0$)

\begin{equation}
\overset{\frown }{\sigma }_{23}^{\mathrm{inc}}\left( \zeta _{1}\right) =%
\mathrm{i}u_{0}k_{T}\mu _{0}\cos \theta _{0}\mathrm{e}^{\mathrm{i}k_{T}\zeta
_{1}\sin \theta _{0}}.  \label{11}
\end{equation}

Par suite, pour respecter l'hypoth\`{e}se de p\'{e}riodicit\'{e} dans la
direction $y_{1}$, nous pouvons \'{e}crire le d\'{e}placement $%
u_{3}^{j}\left( \zeta _{1},0^{+}\right) $ sur la forme \cite{Bloch}~:

\begin{equation}
u_{2}^{j}\left( \zeta _{1},0^{+}\right) =\overset{\frown }{u}_{2}\left(
\zeta _{1}\right) \mathrm{e}^{\mathrm{i}k_{T}jd\sin \theta _{0}},  \label{12}
\end{equation}%
o\`{u} $\overset{\frown }{u}_{2}\left( \zeta _{1}\right) $ ($\left\vert
\zeta _{1}\right\vert <a$) est le \textit{d\'{e}placement diffus\'{e} sur la
fissure de r\'{e}f\'{e}rence} et $j=0,\pm 1,...$. De cette fa\c{c}on, nous
pouvons \'{e}liminer un grand nombre d'inconnues de l'\'{E}q. \eqref{9} en
ne conservant que le d\'{e}placement diffus\'{e} sur la fissure de r\'{e}f%
\'{e}rence.

En substituant \eqref{12} dans \eqref{9} nous obtenons~:

\begin{equation}
u_{2}^{\mathrm{dif}}\left( y_{1},y_{3}\right) =\frac{1}{2\pi }\mathrm{sgn}%
\left( y_{3}\right) \sum_{j=-\infty }^{\infty }\int_{-\infty }^{\infty
}\int_{-a}^{a}\overset{\frown }{u}_{2}\left( \zeta _{1}\right) \mathrm{e}^{%
\mathrm{i}\left( k_{T}\sin \theta _{0}+\xi \right) jd}\mathrm{e}^{\mathrm{i}%
\xi \left( \zeta _{1}-y_{1}\right) }\mathrm{e}^{-\beta \left( \xi \right)
\left\vert y_{3}\right\vert }\mathrm{d}\zeta _{1}\mathrm{d}\xi .  \label{13}
\end{equation}

Rappelons maintenant les propri\'{e}t\'{e}s de la fonction de Dirac $\mathbf{%
\delta }$~:

\begin{equation}
2\pi \sum_{j=-\infty }^{\infty }\mathbf{\delta }\left[ \left( k_{T}\sin \phi
+\xi \right) d-2j\pi \right] =\sum_{j=-\infty }^{\infty }\mathrm{e}^{\mathrm{%
i}j\left( k_{T}\sin \phi +\xi \right) d}~;  \label{14}
\end{equation}

\begin{equation}
\mathbf{\delta }\left( x\right) =\left\vert d\right\vert \mathbf{\delta }%
\left( d\,x\right) ,\text{ \ }\forall x,\text{ \ }\forall d=const\in {%
\mathbb{%
\mathbb{C}
}}~;  \label{15}
\end{equation}

\begin{equation}
\int_{-\infty }^{\infty }\mathbf{\delta }\left( \xi -\xi _{0}\right) \varphi
\left( \xi \right) \mathrm{d}\xi =\varphi \left( \xi _{0}\right) ,\text{ \ }%
\forall \varphi \in C^{1}\left( {\mathbb{%
\mathbb{C}
}}\right) .  \label{16}
\end{equation}

Nous obtenons alors, compte tenu de l'\'{E}q. \eqref{14}~:

\begin{equation}
u_{2}^{\mathrm{dif}}\left( y_{1},y_{3}\right) =\mathrm{sgn}\left(
y_{3}\right) \sum_{j=-\infty }^{\infty }\int_{-\infty }^{\infty
}\int_{-a}^{a}\overset{\frown }{u}_{2}\left( \zeta _{1}\right) \mathbf{%
\delta }\left[ \left( k_{T}\sin \theta _{0}+\xi \right) d-2j\pi \right] 
\mathrm{e}^{\mathrm{i}\xi \left( \zeta _{1}-y_{1}\right) }\mathrm{e}^{-\beta
\left( \xi \right) \left\vert y_{3}\right\vert }\mathrm{d}\zeta _{1}\mathrm{d%
}\xi .  \label{17}
\end{equation}%
Soit, en utilisant l'\'{E}q. \eqref{15}~:

\begin{equation}
u_{2}^{\mathrm{dif}}\left( y_{1},y_{3}\right) =\frac{1}{d}\mathrm{sgn}\left(
y_{3}\right) \sum_{j=-\infty }^{\infty }\int_{-\infty }^{\infty }\mathbf{%
\delta }\left( \xi -\alpha _{j}\right) \varphi \left( \xi \right) \mathrm{d}%
\xi ,  \label{18}
\end{equation}%
avec

\begin{equation}
\varphi \left( \xi \right) =\int_{-a}^{a}\overset{\frown }{u}_{2}\left(
\zeta _{1}\right) \mathrm{e}^{\mathrm{i}\xi \left( \zeta _{1}-y_{1}\right) }%
\mathrm{e}^{-\beta \left( \xi \right) \left\vert y_{3}\right\vert }\mathrm{d}%
\zeta _{1}  \label{19}
\end{equation}%
et

\begin{equation}
\alpha _{j}d=2j\pi -k_{T}d\sin \theta _{0}.  \label{20}
\end{equation}%
Enfin, il r\'{e}sulte des \'{E}qs. \eqref{18}-\eqref{20} apr\`{e}s avoir
utilis\'{e} l'\'{E}q. \eqref{16}, que le champ de d\'{e}placement diffus\'{e}
par l'ensemble des fissures est de la forme~:

\begin{equation}
u_{2}^{\mathrm{dif}}\left( y_{1},y_{3}\right) =\frac{1}{d}\mathrm{sgn}\left(
y_{3}\right) \int_{-a}^{a}\overset{\frown }{u}_{2}\left( \zeta _{1}\right)
S\left( \zeta _{1}-y_{1},y_{3}\right) \mathrm{d}\zeta _{1},  \label{21}
\end{equation}%
avec

\begin{equation}
S\left( x,y\right) =\sum_{j=-\infty }^{\infty }\mathrm{e}^{\mathrm{i}\alpha
_{j}x-\beta _{j}\left\vert y\right\vert },  \label{22}
\end{equation}%
o\`{u}

\begin{equation}
\beta _{j}=\beta \left( \alpha _{j}\right) .  \label{23}
\end{equation}%
L'\'{e}quation \eqref{21} montre qu'il est possible d'\'{e}valuer $u_{2}^{%
\mathrm{dif}}$ en tous points de l'espace, si le champ diffus\'{e} $\overset{%
\frown }{u}_{2}\left( \zeta _{1}\right) $ est connu pour $\left\vert \zeta
_{1}\right\vert <a$. Par cons\'{e}quent, $\overset{\frown }{u}_{2}$ sert de 
\textit{quantit\'{e} inconnue fondamentale}.

Le champ de contrainte associ\'{e} au champ de d\'{e}placement \eqref{21} s'%
\'{e}crit~:

\begin{equation}
\sigma _{23}^{\mathrm{dif}}\left( y_{1},y_{3}\right) =\frac{\mu _{0}}{d}%
\int_{-a}^{a}\overset{\frown }{u}_{2}\left( \zeta _{1}\right) S^{\prime
}\left( \zeta _{1}-y_{1},y_{3}\right) \mathrm{d}\zeta _{1},  \label{24}
\end{equation}%
avec

\begin{equation}
S^{\prime }\left( x,y\right) =-\sum_{j=-\infty }^{\infty }\beta _{j}\mathrm{e%
}^{\mathrm{i}\alpha _{j}x-\beta _{j}\left\vert y\right\vert }.  \label{25}
\end{equation}

En faisant s'approcher le point d'observation $\left( y_{1},y_{3}\right) $
des l\`{e}vres de la fissure de r\'{e}f\'{e}rence dans l'\'{E}q. \eqref{24},
et en utilisant la condition aux limites \eqref{4}, nous obtenons l'\'{e}%
quation int\'{e}grale suivante pour l'inconnue $\overset{\frown }{u}_{2}$~:

\begin{equation}
\int_{-a}^{a}\overset{\frown }{u}_{2}\left( \zeta _{1}\right)
\sum_{j=-\infty }^{\infty }\beta _{j}\mathrm{e}^{\mathrm{i}\alpha _{j}\left(
\zeta _{1}-y_{1}\right) }\mathrm{d}\zeta _{1}=-\mathrm{i}u_{0}k_{T}d\cos
\theta _{0}\mathrm{e}^{\mathrm{i}k_{T}y_{1}\sin \theta _{0}},\text{ \ }%
\left\vert y_{1}\right\vert <a.  \label{26}
\end{equation}%
L'\'{e}quation int\'{e}grale \eqref{26} peut \^{e}tre r\'{e}solue num\'{e}%
riquement en adoptant une m\'{e}thode dite de Galerkin (voir \cite{ZG} pour
plus de d\'{e}tails). La solution $\overset{\frown }{u}_{2}$ de l'\'{e}%
quation int\'{e}grale \eqref{26} est ainsi d\'{e}termin\'{e}e par un calcul
exact de diffusion multiple prenant en compte toutes les interactions
possibles entre les fissures.

\subsubsection{ Coefficients de r\'{e}flexion et de transmission du mode
fondamental}

Dans l'\'{E}q. \eqref{21}, chaque $j$ correspond \`{a} un mode d'onde. Ceci
implique que le champ de d\'{e}placement diffus\'{e} consiste en un nombre
infini de modes, chacun avec sa propre fr\'{e}quence de coupure. Pour $%
\alpha _{j}^{2}>k_{T}^{2}$ o\`{u}~:

\begin{equation}
\left\vert \frac{2j\pi }{d}-k_{T}\sin \theta _{0}\right\vert >k_{T},
\label{27}
\end{equation}%
les modes diffus\'{e}s de $u_{2}^{\mathrm{dif}}$ sont des ondes
stationnaires (ou \'{e}vanescentes) qui d\'{e}croissent exponentiellement
avec l'augmentation de la distance du r\'{e}seau, tandis que pour $\alpha
_{j}^{2}>k_{T}^{2}$, ces modes repr\'{e}sentent des ondes homog\`{e}nes se
propageant dans les directions $y_{3}$ positives (ou n\'{e}gatives). Aux fr%
\'{e}quences de coupure $\alpha _{j}^{2}=k_{T}^{2}$, un mode stationnaire se
converti dans un mode homog\`{e}ne. On peut montrer que, suffisamment loin
du r\'{e}seau et \`{a} des fr\'{e}quences suffisamment basses, le mode
d'ordre z\'{e}ro ($j=0$) sera dominant. Ces ondes d'ordre z\'{e}ro sont les 
\textit{ondes planes homog\`{e}nes r\'{e}fl\'{e}chies et transmises}, qui se
propagent dans la m\^{e}me direction que l'onde incidente. Ceci devient
clair en r\'{e}\'{e}crivant le mode d'ordre z\'{e}ro sous la forme~:

\begin{equation}
u_{3}^{\mathrm{dif}}\left( y_{1},y_{3}\right) =R_{0}\mathrm{e}^{\mathrm{i}%
k_{T}\left( y_{1}\sin \theta _{0}-y_{3}\cos \theta _{0}\right) },\text{ \ }%
y_{3}<0,  \label{28}
\end{equation}

\begin{equation}
u_{3}^{\mathrm{dif}}\left( y_{1},y_{3}\right) =\left( T_{0}-1\right) \mathrm{%
e}^{\mathrm{i}k_{T}\left( y_{1}\sin \theta _{0}+y_{3}\cos \theta _{0}\right)
},\text{ \ }y_{3}>0,  \label{29}
\end{equation}%
o\`{u} $R_{0}$ et $T_{0}$ d\'{e}signent les coefficients de r\'{e}flexion et
de transmission du \textit{mode fondamental} (d'ordre z\'{e}ro) qui sont donn%
\'{e}s par~:

\begin{equation}
R_{0}=-\frac{1}{d}\int_{-a}^{a}\overset{\frown }{u}_{2}\left( v\right) 
\mathrm{e}^{-\mathrm{i}k_{T}v\sin \theta _{0}}\mathrm{d}v,\text{ \ }%
T_{0}=1-R_{0},  \label{30}
\end{equation}%
Notons que l'\'{E}q. \eqref{30} satisfait le bilan d'\'{e}nergie bien connu 
\cite{Ali}~:

\begin{equation}
\left\vert R_{0}\right\vert ^{2}+\left\vert T_{0}\right\vert ^{2}=1.
\label{31}
\end{equation}

\subsection{ R\'{e}sultats num\'{e}riques et discussion}

Le coefficient de r\'{e}flexion du mode fondamental est trac\'{e} ici sur
les Figs. 2-4 comme \'{e}tant une fonction de l'espacement entre fissures $%
\tilde{d}=d/a$, de la fr\'{e}quence d'excitation $\tilde{\omega }=k_{T} a$
et de l'angle d'incidence $\theta _{0} $.

La Fig. 2 montre le module du coefficient de r\'{e}flexion $R_{0}$ en
fonction de la fr\'{e}quence adimensionnelle $\tilde{\omega}$ pour trois
valeurs de $\tilde{d}$ et pour deux angles d'incidence ($\theta _{0}=0$ et $%
\theta _{0}=\pi /4$). Le coefficient de r\'{e}flexion est z\'{e}ro \`{a} $%
\tilde{\omega}=0$ (ondes infiniment longues), et augmente monotonement avec
l'augmentation de $\tilde{\omega}$, jusqu'\`{a} ce qu'un maximum soit
atteint et pour lequel les courbes affichent une pente discontinue. Cette
discontinuit\'{e} ainsi que les suivantes qui apparaissent lorsque $\tilde{%
\omega}$ augmente, se produisent aux fr\'{e}quences de coupure $\tilde{\omega%
}^{\left( j\right) }$, ce qui r\'{e}sulte des \'{E}qs. \eqref{B}, \eqref{20}
et \eqref{23},

\begin{equation}
\tilde{\omega}^{\left( j\right) }=\frac{2j\pi }{\tilde{d}}-\tilde{\omega}%
^{\left( j\right) }\sin \theta _{0},\text{ \ }j=1,2,....  \label{32}
\end{equation}%
Ce sont les valeurs de $\tilde{\omega}$ auxquelles un mode d'ordre sup\'{e}%
rieur devient un mode propagatif (homog\`{e}ne). Pour l'incidence normale ($%
\theta _{0}=0$), l'\'{E}q. \eqref{32} se r\'{e}duit \`{a} $\tilde{\omega}%
^{\left( j\right) }=2j\pi /\tilde{d}$. Consid\'{e}rons par exemple la courbe
pour $\tilde{d}=2$ et $\theta _{0}=0$. Dans la gamme de fr\'{e}quences $%
0\leq \tilde{\omega}<\pi $, le champ r\'{e}fl\'{e}chi \`{a} une distance
quelconque de $y_{3}=0$ est uniquement constitu\'{e} du terme $j=0$ dans l'%
\'{E}q. \eqref{22}~; les termes d'ordres sup\'{e}rieurs peuvent \^{e}tre n%
\'{e}glig\'{e}s, car ils sont \'{e}vanescents. Dans les gammes de fr\'{e}%
quences $\pi <\tilde{\omega}<2\pi $ et $2\pi <\tilde{\omega}<4\pi $, les
termes correspondant au $j=1$ et $j=2$ doivent, pourtant, aussi \^{e}tre
inclus dans le champ total r\'{e}fl\'{e}chi en champ lointain. Seul le
coefficient de r\'{e}flexion $\left\vert R_{0}\right\vert $ du mode
fondamental est repr\'{e}sent\'{e} sur la Fig. 2. Pour une valeur
approximative $\tilde{d}\simeq 2$ (fissures \'{e}troitement align\'{e}es),
le coefficient de r\'{e}flexion approche rapidement $\left\vert
R_{0}\right\vert =1$ (r\'{e}flexion totale) lorsque $\tilde{\omega}$
augmente. Les courbes de la Fig. 2b ($\theta _{0}=\pi /4$) ont la m\^{e}me
apparence g\'{e}n\'{e}rale que celles de la Fig. 2a ($\theta _{0}=0$).
Notons pourtant que, pour les m\^{e}mes valeurs de $\tilde{d}$ et $\tilde{%
\omega}$, les coefficients de r\'{e}flexion semblent \^{e}tre plus petits
pour $\theta _{0}=\pi /4$ que pour $\theta _{0}=0$, du moins, dans la gamme
de fr\'{e}quences $0\leq \tilde{\omega}\leq \pi /2$.

Pour $\tilde{d}=20$, l'espacement entre fissures est suffisamment grand afin
pour que l'on n\'{e}glige les interactions entre les fissures, et les
coefficients de r\'{e}flexion montr\'{e}s sur la Fig. 3 approchent ceux
d'une fissure seule~:

\begin{equation}
\left\vert R_{0}\right\vert \simeq \frac{\left\vert f\left( \pi -\theta
_{0},\theta _{0}\right) \right\vert }{\pi \tilde{\omega}\cos \theta _{0}},%
\text{ \ }(\tilde{d}\gg 1)~;  \label{33)}
\end{equation}%
$f\left( \theta ,\theta _{0}\right) $ \'{e}tant l'amplitude de diffusion en
champ lointain de la fissure, \'{E}q. \eqref{Ftheta}. Notons que $f\left(
\pi -\theta _{0},\theta _{0}\right) =-f\left( \theta _{0},\theta _{0}\right) 
$, o\`{u} $f\left( \theta _{0},\theta _{0}\right) $ repr\'{e}sente
l'amplitude de diffusion vers l'avant. Ceci nous fournit un contr\^{o}le sur
l'exactitude de la m\'{e}thode num\'{e}rique pr\'{e}sent\'{e}e ici.

Dans la Fig. 4, la variation du coefficient de r\'{e}flexion en fonction de
la distance adimensionnelle $\tilde{d}$ est montr\'{e}e pour une incidence
normale ($\theta _{0}=0$), et pour plusieurs valeurs de la fr\'{e}quence
adimensionnelle $\tilde{\omega}$. Lorsque $\tilde{d}$ s'approche de $2$
(fissures \'{e}troitement align\'{e}es), nous trouvons que $\left\vert
R_{0}\right\vert $ tend vers $1$ (r\'{e}flexion totale). Remarquons que sur
la courbe correspondant \`{a} $\tilde{\omega}=2$, un pic se distingue
lorsque $\tilde{d}=\pi $. En effet, au regard de l'\'{E}q. \eqref{32}, $%
\tilde{\omega}=2$ repr\'{e}sente une fr\'{e}quence de coupure pour une
distance $\tilde{d}=\pi $.

\section{ R\'{e}seau p\'{e}riodique bidimensionnel de fissures}

La propagation d'ondes antiplanes dans des r\'{e}seaux bidimensionnels de
fissures est \'{e}tudi\'{e}e. Ces r\'{e}seaux sont form\'{e}s d'un
empilement de r\'{e}seaux lin\'{e}aires infinis et p\'{e}riodiques. Dans un
premier temps, nous rappelons la m\'{e}thode de calcul des coefficients de r%
\'{e}flexion et de transmission de r\'{e}seaux bidimensionnels finis et p%
\'{e}riodiques. Ensuite, pour des r\'{e}seaux infinis et p\'{e}riodiques, le
nombre d'onde des ondes antiplanes se propageant dans le milieu homog\`{e}ne 
\'{e}quivalent au r\'{e}seau bidimensionnel de fissures est calcul\'{e} \`{a}
l'aide de la th\'{e}orie de Floquet.

\subsection{ Coefficients de r\'{e}flexion et de transmission d'un r\'{e}%
seau fini}

Consid\'{e}rons le r\'{e}seau global de la Fig. 5. Il est compos\'{e} d'un
nombre arbitraire $S$ de r\'{e}seaux lin\'{e}aires infinis, r\'{e}guli\`{e}%
rement espac\'{e}s d'une distance $D$ le long de l'axe $y_{3}$. Le r\'{e}%
seau global est donc fini et d'\'{e}paisseur $h=\left( S-1\right) D$. Nous
consid\'{e}rons tout d'abord que tous les r\'{e}seaux plans sont identiques, 
$d$ \'{e}tant la p\'{e}riode spatiale commune (selon l'axe $y_{1}$) et $2a$
la taille des fissures. Isolons ici un r\'{e}seau plan $s$ ($s=1,2,...,S$)
du r\'{e}seau global. Le probl\`{e}me de diffusion d'une onde antiplane par
un tel r\'{e}seau a \'{e}t\'{e} \'{e}tudi\'{e} dans la section pr\'{e}c\'{e}%
dente. Les coefficients de r\'{e}flexion et de transmission du r\'{e}seau lin%
\'{e}aire infini $s$ sont d\'{e}sormais connus et not\'{e}s $R$ et $T$, o%
\`{u} l'indice $0$ associ\'{e} au mode fondamental a \'{e}t\'{e} ici supprim%
\'{e} pour all\'{e}ger les \'{e}critures. \`{A} partir de ces coefficients,
ceux du r\'{e}seaux global, $\mathcal{R}$ et $\mathcal{T}$, peuvent \^{e}tre
calcul\'{e}s en \'{e}laborant un sch\'{e}ma it\'{e}ratif relativement
simple. La technique employ\'{e}e se fonde sur celle de Huang et Heckl \cite%
{HH}. Ce formalisme est valable uniquement aux basses fr\'{e}quences et
seuls les modes d'ordre z\'{e}ro peuvent se propager dans le r\'{e}seau. Les
autres modes sont des modes \'{e}vanescents qui d\'{e}croissent
exponentiellement au fur et \`{a} mesure de leur propagation.

Dans la suite, nous mettons en \'{e}quation les r\'{e}flexions multiples et
les allers-retours entre deux r\'{e}seaux lin\'{e}aires cons\'{e}cutifs. Le
calcul des coefficients de r\'{e}flexion et de transmission du r\'{e}seau
bidimensionnel s'effectue sans difficult\'{e} en utilisant un algorithme it%
\'{e}ratif pr\'{e}sent\'{e} par la suite.

\subsubsection{ R\'{e}flexion et transmission par deux r\'{e}seaux lin\'{e}%
aires infinis}

Isolons ici deux r\'{e}seaux lin\'{e}aires parall\`{e}les cons\'{e}cutifs $r$
et $s$ du r\'{e}seau global, le r\'{e}seau $r$ \'{e}tant celui soumis \`{a}
l'onde incidente.

Le champ total transmis, compte tenu des r\'{e}flexions multiples entre les r%
\'{e}seaux $r$ et $s$, s'\'{e}crit \`{a} l'aide des s\'{e}ries de Debye,
sous la forme suivante~:

\begin{equation}
\begin{array}{l}
{u_{2}^{T}=u_{0}T_{r}T_{s}+u_{0}T_{r}T_{s}R_{s}R_{r}\mathrm{e}^{\mathrm{i}%
\phi }+u_{0}T_{r}T_{s}\left( R_{s}R_{r}\mathrm{e}^{\mathrm{i}\phi }\right)
^{2}+...} \\ 
{=u_{0}T_{r}T_{s}\left( 1+R_{s}R_{r}\mathrm{e}^{\mathrm{i}\phi }+\left(
R_{s}R_{r}\mathrm{e}^{\mathrm{i}\phi }\right) ^{2}+...\right) } \\ 
{=u_{0}T_{r}T_{s}\left( 1-R_{s}R_{r}\mathrm{e}^{\mathrm{i}\phi }\right)
^{-1}.}%
\end{array}
\label{ZEqnNum974604}
\end{equation}%
Le champ total transmis a \'{e}t\'{e} exprim\'{e} en fonction d'une s\'{e}%
rie g\'{e}om\'{e}trique. La convergence de cette s\'{e}rie est assur\'{e}e d%
\`{e}s lors que $\left\vert R_{s}R_{r}\mathrm{e}^{\mathrm{i}\phi
}\right\vert <1$. Dans l'\'{E}q. \eqref{ZEqnNum974604}, nous avons introduit
la notation~:

\begin{equation}  \label{35)}
\phi =2k_{T} D\cos \theta _{0} .
\end{equation}

D'une mani\`{e}re similaire, le champ total r\'{e}fl\'{e}chi par les deux r%
\'{e}seaux lin\'{e}aires peut \^{e}tre obtenu sous la forme~:

\begin{equation}
u_{2}^{R}=u_{0}+u_{0}R_{r}+u_{0}T_{r}^{2}R_{s}\mathrm{e}^{\mathrm{i}\phi
}\left( 1-R_{s}R_{r}\mathrm{e}^{\mathrm{i}\phi }\right) ^{-1}.  \label{36)}
\end{equation}

Les coefficients de r\'{e}flexion et de transmission globaux sont alors d%
\'{e}finis par~:

\begin{equation}
\mathcal{R}_{2}\equiv \frac{u_{2}^{R}}{u_{0}}-1=R_{r}+\frac{T_{r}^{2}R_{s}%
\mathrm{e}^{\mathrm{i}\phi }}{1-R_{s}R_{r}\mathrm{e}^{\mathrm{i}\phi }},
\label{ZEqnNum843153}
\end{equation}

\begin{equation}
\mathcal{T}_{2}\equiv \frac{u_{2}^{T}}{u_{0}}=\frac{T_{r}T_{s}}{1-R_{s}R_{r}%
\mathrm{e}^{\mathrm{i}\phi }}.  \label{ZEqnNum706012}
\end{equation}

\subsubsection{ Coefficients de r\'{e}flexion et de transmission du r\'{e}%
seau global}

Les formules \eqref{ZEqnNum843153} et \eqref{ZEqnNum706012} peuvent \^{e}tre
utilis\'{e}es pour \'{e}laborer une m\'{e}thode it\'{e}rative de calcul des
coefficients de r\'{e}flexion et de transmission. Nous d\'{e}composons le r%
\'{e}seau bidimensionnel dans $S-1$ r\'{e}seaux (ceux-ci forment alors un 
\guillemotleft ~sous-r\'{e}seau~\guillemotright\ bidimensionnel caract\'{e}%
ris\'{e} par les coefficients inconnus $\mathcal{R}_{S-1}$ et $\mathcal{T}%
_{S-1}$) et dans un autre r\'{e}seau lin\'{e}aire (caract\'{e}ris\'{e} par
les coefficients locaux $R$ et $T$). Suite \`{a} cette d\'{e}composition,
les coefficients $\mathcal{R}_{S}$ et $\mathcal{T}_{S}$, du r\'{e}seau
global de la Fig. 5, peuvent s'exprimer en fonction de ceux du sous-r\'{e}%
seau $S-1$, en rempla\c{c}ant $\left( R_{r},T_{r}\right) $ et $\left(
R_{s},T_{s}\right) $ dans les \'{E}qs. \eqref{ZEqnNum843153}-%
\eqref{ZEqnNum706012} par $\left( R,T\right) $ et $\left( \mathcal{R}_{S-1},%
\mathcal{T}_{S-1}\right) $, respectivement. Ainsi, nous construisons les
formules it\'{e}ratives suivantes~:

\begin{equation}
\mathcal{R}_{S}=R+\frac{T^{2}\mathcal{R}_{S-1}\mathrm{e}^{\mathrm{i}\phi }}{%
1-R\,\mathcal{R}_{S-1}\mathrm{e}^{\mathrm{i}\phi }},  \label{39)}
\end{equation}

\begin{equation}
\mathcal{T}_{S}=\frac{T\mathcal{T}_{S-1}}{1-R\,\mathcal{R}_{S-1}\mathrm{e}^{%
\mathrm{i}\phi }}.  \label{40)}
\end{equation}%
Notons que les coefficients de r\'{e}flexion et de transmission calcul\'{e}s
peuvent \^{e}tre valid\'{e}s en utilisant directement la relation de
conservation d'\'{e}nergie \eqref{31}. Il suffit de remplacer dans cette
relation les coefficients du r\'{e}seau lin\'{e}aire par ceux du r\'{e}seau
bidimensionnel.

\subsubsection{ Exemples num\'{e}riques}

Les r\'{e}seaux bidimensionnels consid\'{e}r\'{e}s dans ce paragraphe poss%
\`{e}dent une direction de p\'{e}riodicit\'{e} selon l'axe $y_{1}$.\ Pour
simplifier l'analyse des r\'{e}sultats, les coefficients de r\'{e}flexion
(transmission) sont ici calcul\'{e}s en incidence normale. Deux exemples num%
\'{e}riques sont consid\'{e}r\'{e}s : i) une r\'{e}partition p\'{e}riodique
selon l'axe $y_{3}$ ; ii) une r\'{e}partition al\'{e}atoire de r\'{e}seaux
lin\'{e}aires selon l'axe $y_{3}$.

\paragraph{R\'{e}partition p\'{e}riodique de r\'{e}seaux lin\'{e}aires.}

Des r\'{e}sultats num\'{e}riques sont obtenus dans le cas d'un r\'{e}seau
bidimensionnel compos\'{e} de $S=1$, $2$, $5$ et $15$ r\'{e}seaux lin\'{e}%
aires infinis. La Fig. 6 montre l'influence du nombre des r\'{e}seaux lin%
\'{e}aires $S$ sur le coefficient de transmission en \'{e}nergie $\left\vert 
\mathcal{T}\right\vert ^{2}$. Pour $\tilde{d}(=d/a)=5$ et $\tilde{D}(=D/d)=1$%
, la premi\`{e}re fr\'{e}quence de coupure correspond alors \`{a} $%
k_{T}D=2\pi $ (voir l'\'{E}q. \eqref{32}). Observons que dans une gamme de
basses fr\'{e}quences, le nombre des minima et des maxima \'{e}gale le
nombre des rang\'{e}es des r\'{e}seaux lin\'{e}aires moins un. La
comparaison des courbes pour $S=5$ et $S=15$ illustre la formation des
bandes passantes et bandes d'arr\^{e}t (ou interdites). L'existence des
bandes passantes et d'arr\^{e}t donnent au r\'{e}seau bidimensionnel les
caract\'{e}ristiques d'un filtre m\'{e}canique \cite{Braga} dans lequel des
ondes ayant la fr\'{e}quence et le nombre d'onde se trouvant dans les bandes
d'arr\^{e}t sont att\'{e}nu\'{e}es, alors que d'autres sont transmises. M%
\^{e}me si le nombre des r\'{e}seaux lin\'{e}aires est fini, de caract\'{e}%
ristiques attendues pour un milieu p\'{e}riodique infini, telles que des
bandes passantes et d'arr\^{e}t sont reconnues. Ces aspects sont \'{e}tudi%
\'{e}s en d\'{e}tail dans la section suivante.

\paragraph{R\'{e}partition al\'{e}atoire de r\'{e}seaux lin\'{e}aires.}

Consid\'{e}rons ici un r\'{e}seau bidimensionnel compos\'{e} de $S=30$ r\'{e}%
seaux lin\'{e}aires infinis et dispos\'{e}s p\'{e}riodiquement avec une p%
\'{e}riode $\tilde{D}=1$. Chaque r\'{e}seau lin\'{e}aire est caract\'{e}ris%
\'{e} par $\tilde{d}=5$. L'objectif de ce paragraphe est d'identifier
l'effet d'un d\'{e}faut de position al\'{e}atoire de chacun des r\'{e}seaux
lin\'{e}aires par rapport \`{a} leur position initiale p\'{e}riodique. Seuls
les premiers et derniers r\'{e}seaux lin\'{e}aires sont suppos\'{e}es fixes.
Nous introduisons une distance limite d'approche entre les r\'{e}seaux, not%
\'{e}e $b$, dans le but de quantifier ce d\'{e}faut de p\'{e}riodicit\'{e}.
Ainsi, $b=D$ correspond \`{a} un r\'{e}seau p\'{e}riodique et, \`{a} l'extr%
\^{e}me, $b=0$ \`{a} un r\'{e}seau ap\'{e}riodique o\`{u} les fissures de
deux r\'{e}seaux lin\'{e}aires cons\'{e}cutifs sont susceptibles d'\^{e}tre
en contact. Nous nous assurons ici que $b>4D$ pour que la contribution
d'ondes \'{e}vanescentes issues d'un r\'{e}seau lin\'{e}aire soit n\'{e}%
gligeable vue des r\'{e}seaux voisins.

En fixant une valeur de $b$, la distribution des r\'{e}seaux lin\'{e}aires
se r\'{e}sume \`{a} calculer le nouvel espacement le long de l'axe $y_{3}$,

\begin{equation*}
D_{s}=D+(0,5-\delta _{j})(D-b),\ \ \ (s=2,3,...,S-1),
\end{equation*}%
o\`{u} $\delta _{j}\in \lbrack 0,1]$ est une variable al\'{e}atoire ob\'{e}%
issant une loi de probabilit\'{e} uniforme. Cette variable est propre \`{a}
chaque r\'{e}seau lin\'{e}aire $S$. A titre d'exemple, la Fig. 7 illustre la
r\'{e}partition des r\'{e}seaux lin\'{e}aires pour $b=0,9D$ et $b=0,6D$ et
pour une s\'{e}rie de tirages al\'{e}atoires $j$ de $1$ \`{a} $10$. Pour une
bonne lisibilit\'{e} de la repr\'{e}sentation seulement 11 r\'{e}seaux sont
illustr\'{e}s. Lorsque $b=0,9D$, le d\'{e}placement d'un r\'{e}seau lin\'{e}%
aire autour de sa position initiale est inf\'{e}rieur \`{a} $0,5a$, tandis
que le cas $b=0,6D$ correspond \`{a} un d\'{e}placement des r\'{e}seaux lin%
\'{e}aires inferieur \`{a} $2a$ autour de leur position initiale.

La Fig. 8 pr\'{e}sente les coefficients de r\'{e}flexion en \'{e}nergie $%
\left\vert \mathcal{R}\right\vert ^{2}$ obtenus pour deux distances limites
d'approche : $b=0,9D$ et $b=0,6D$. Ces courbes sont compar\'{e}es avec celle
obtenue pour $b=D$, c'est-\`{a}-dire en situation p\'{e}riodique. Nous avons
adopt\'{e} une \'{e}chelle logarithmique pour $\left\vert \mathcal{R}%
\right\vert ^{2}$ afin de mieux juger cette comparaison. Le cas $b=0,9a$
correspond \`{a} un faible d\'{e}sordre par rapport \`{a} la situation p\'{e}%
riodique. Pour ce faible d\'{e}sordre, la pr\'{e}sence de bandes passantes
et d'arr\^{e}t est toujours observable. Lorsque l'on diminue la distance
limite d'approche, $b=0,6D$, la situation physique est diff\'{e}rente. M\^{e}%
me si la p\'{e}riodicit\'{e} est globalement pr\'{e}serv\'{e}e, la Fig. 8
affiche une diminution des bandes passantes.

\subsection{ Dispersion et att\'{e}nuation des ondes se propageant dans le r%
\'{e}seau infini}

Le r\'{e}seau global de la Fig. 5 est maintenant suppos\'{e} \^{e}tre
constitu\'{e} d'un nombre infini de r\'{e}seaux lin\'{e}aires infinis. Ces
derniers sont tous identiques et r\'{e}partis r\'{e}guli\`{e}rement avec une
p\'{e}riode $D$ le long de l'axe $y_{3}$. Les fissures de taille $2a$ ont
leur centre localis\'{e} aux positions $y_{1}=jd$, $y_{3}=\ell D$ ($j,\ell
=0,\pm 1,...$). Il est bien connu de la th\'{e}orie de Floquet (p\'{e}%
riodicit\'{e} 1D) ou de Bloch (p\'{e}riodicit\'{e} 2D et 3D) \cite{Ince,
Odeh} que les \'{e}quations diff\'{e}rentielles avec des coefficients p\'{e}%
riodiques admettent des solutions de la forme~:

\begin{equation}
u_{2}\left( y_{3}\right) =w\left( y_{3}\right) \mathrm{e}^{\mathrm{i}\mathit{%
\gamma }y_{3}}\text{, \ o\`{u} \ }w\left( y_{3}\right) =w\left(
y_{3}+D\right) .  \label{41}
\end{equation}%
La grandeur $\mathit{\gamma }$ a le caract\`{e}re d'un nombre d'onde
effectif, caract\'{e}risant la propagation des ondes dans la direction $%
y_{3} $, dans un milieu \'{e}lastique homog\`{e}ne \'{e}quivalent au r\'{e}%
seau bidimensionnel p\'{e}riodique infini. On dit que $\mathit{\gamma }$ est
le \textit{nombre d'onde de Floquet}. Ce nombre d'onde est \`{a} valeurs
complexes et peut s'\'{e}crire~:

\begin{equation}
\mathit{\gamma }=\mathit{\gamma }^{\prime }+\mathrm{i}\mathit{\gamma }%
^{\prime \prime },  \label{42)}
\end{equation}%
o\`{u} la partie r\'{e}elle $\mathit{\gamma }^{\prime }=\omega /c^{\prime }$
est la composante suivant $y_{3}$ du vecteur d'onde $\vec{\mathit{\gamma }}$
des ondes se propageant avec la vitesse $c^{\prime }$ dans ledit milieu
effectif, et la partie imaginaire $\mathit{\gamma }^{\prime \prime }$ est
leur att\'{e}nuation.

Il est relativement facile de calculer $\mathit{\gamma }$.\ Achenbach et Li 
\cite{Ali} ont pr\'{e}sent\'{e} une m\'{e}thode de calcul du nombre d'onde $%
\mathit{\gamma }$ (voir aussi \cite{AA}). La d\'{e}rivation formelle de l'%
\'{e}quation de dispersion est bas\'{e}e sur la p\'{e}riodicit\'{e} du
solide dans la direction $y_{3}$. Une \'{e}quation de dispersion qui relie $%
\mathit{\gamma }D$ et $k_{T}D$ peut s'\'{e}crire sous la forme \cite{Ali} :

\begin{equation}
2T\cos \left( \mathit{\gamma }D\right) =\left( T^{2}-R^{2}-1\right) \mathrm{e%
}^{\mathrm{i}k_{T}D}+2\cos \left( k_{T}D\right) ,  \label{ZEqnNum710474}
\end{equation}%
o\`{u} $T$ et $R$ sont les coefficients de r\'{e}flexion et de transmission
du mode fondamental d'un r\'{e}seau lin\'{e}aire infini.

L'existence de solutions r\'{e}elles et complexes du nombre d'onde de
Floquet peut \^{e}tre interpr\'{e}t\'{e}e en termes desdites bandes
passantes et d'arr\^{e}t en fr\'{e}quence. Ce sont, respectivement, des
parties du spectre o\`{u} les ondes harmoniques sont propag\'{e}es ou att%
\'{e}nu\'{e}es, et des caract\'{e}ristiques des ondes se propageant dans des
milieux multicouches [12]. Sur les bandes passantes, le nombre d'onde de
Floquet est r\'{e}el, et les ondes sont propag\'{e}es essentiellement \`{a}
travers le solide non-att\'{e}nu\'{e}es. Les bandes d'arr\^{e}t sont des r%
\'{e}gions o\`{u} le spectre est imaginaire pur ou complexe. Ici, en vertu
de la solution de Floquet \eqref{41}, les ondes sont att\'{e}nu\'{e}es.
Physiquement, l'att\'{e}nuation est due \`{a} la r\'{e}flexion progressive
d'une onde se dirigeant dans le sens positif aux r\'{e}seaux lin\'{e}aires
successifs du r\'{e}seau bidimensionnel.

\subsubsection{ \'{E}tude num\'{e}rique}

Des courbes de dispersion associ\'{e}es au r\'{e}seau infini sont compar\'{e}%
es avec les r\'{e}sultats obtenus pour un r\'{e}seau fini. Cette comparaison
n'est valable que si le r\'{e}seau fini contient suffisamment de r\'{e}seaux
lin\'{e}aires. Avec $S=15$, cette condition est satisfaite. Ceci est justifi%
\'{e} par le fait que les coefficients de transmission (ou de r\'{e}flexion)
r\'{e}v\`{e}lent les m\^{e}mes bandes interdites que les courbes de
dispersion (ou d'att\'{e}nuation) et que celles-ci restent inchang\'{e}es
pour un nombre $S$ plus important de r\'{e}seaux lin\'{e}aires.

Pour $\tilde{d}=2,5$ et $\tilde{D}=4$ (la premi\`{e}re fr\'{e}quence de
coupure correspond alors \`{a} $k_{T}D=4\pi $), la Fig. 9 compare les
courbes de dispersion (et d'att\'{e}nuation) du r\'{e}seau de fissures avec
le coefficient de transmission en \'{e}nergie du r\'{e}seau fini. Les
courbes de la Fig. 9a font appara\^{\i}tre une s\'{e}rie des bandes d'arr%
\^{e}t pour lesquelles $\mathit{\gamma }^{\prime }=0$ ou $\mathit{\gamma }%
^{\prime }=\pi /D$. En termes de cristallographie, $\mathit{\gamma }^{\prime
}=0$ correspond au centre de la premi\`{e}re zone de Brillouin, et $\mathit{%
\gamma }^{\prime }=\pi /D$ \`{a} un vecteur d'onde sur la fronti\`{e}re de
cette zone. Lorsque $\mathit{\gamma }^{\prime }=\pi /D$, un r\'{e}gime
d'ondes stationnaires s'\'{e}tablit. Dans une bande d'arr\^{e}t, les ondes
se propageant dans le r\'{e}seau sont att\'{e}nu\'{e}es. Inversement, dans
les bandes passantes, les ondes sont dispersives et se propagent dans le r%
\'{e}seau sans s'att\'{e}nuer ($\mathit{\gamma }^{\prime \prime }=0$).
Notons que les branches imaginaires du spectre d'att\'{e}nuation doivent
tirer leur origine des points sur les branches r\'{e}elles du spectre de
dispersion o\`{u} la pente est nulle \cite{Onoe}. La premi\`{e}re bande d'arr%
\^{e}t est localis\'{e}e autour d'une fr\'{e}quence telle que la distance $D$
co\"{\i}ncide avec une demi-longueur d'onde de l'onde incidente~: $%
k_{T}D=\pi $. Les bandes d'arr\^{e}t suivantes sont localis\'{e}es \`{a} des
multiples entiers de cette fr\'{e}quence. La relation ci-avant a \'{e}t\'{e} 
\'{e}tablie par Bragg en cristallographie. Il s'agit de la condition d'interf%
\'{e}rence constructive entre deux ondes r\'{e}fl\'{e}chies par deux r\'{e}%
seaux lin\'{e}aires (ou plans r\'{e}ticulaires) distants de $D$. Pour un
grand nombre de plans r\'{e}ticulaires, une grande quantit\'{e} d'ondes
interf\`{e}rent et donnent naissance aux bandes interdites observ\'{e}es.
Les largeurs des bandes d'arr\^{e}t sont li\'{e}es \`{a} la densit\'{e} des
fissures dans les r\'{e}seaux lin\'{e}aires~: plus le rapport $\tilde{d}$
est petit, plus les bandes d'arr\^{e}t s'\'{e}largissent et donc plus les
bandes passantes sont r\'{e}duites. Pour des r\'{e}seaux tr\`{e}s denses,
les bandes passantes disparaissent compl\`{e}tement car chaque r\'{e}seau lin%
\'{e}aire r\'{e}fl\'{e}chit beaucoup plus qu'il ne transmet. Les ondes
transmises sont att\'{e}nu\'{e}es lors du franchissement de chaque r\'{e}%
seau lin\'{e}aire et sont quasi inexistantes apr\`{e}s quelques p\'{e}riodes 
$D$ parcourues dans le r\'{e}seau bidimensionnel.

Dans le m\^{e}me domaine d'\'{e}tude, la Fig. 9b pr\'{e}sente l'\'{e}%
volution du coefficient de transmission en \'{e}nergie $\left\vert \mathcal{T%
}\right\vert ^{2}$ d'un r\'{e}seau fini form\'{e} de $15$ rang\'{e}es des
fissures. \`{A} chaque bande d'arr\^{e}t observ\'{e}e sur le spectre de
dispersion et d'att\'{e}nuation, Fig. 9a, co\"{\i}ncide ici une transmission
nulle. Les bandes interdites sont effectivement d\'{e}sign\'{e}es par le
coefficient de transmission et des calculs men\'{e}s pour un plus grand
nombre de r\'{e}seaux lin\'{e}aires n'ont r\'{e}v\'{e}l\'{e} aucune
modification quant \`{a} leur forme. Ceci confirme que, pour $S=15$, le r%
\'{e}seau global comporte suffisamment de r\'{e}seaux lin\'{e}aires pour que
la propagation d'ondes puisse \^{e}tre d\'{e}crite par la th\'{e}orie de
Floquet. Par rapport aux courbes de dispersion et d'att\'{e}nuation, le trac%
\'{e}e du coefficient de transmission fournit un r\'{e}sultat nouveau li\'{e}
\`{a} la dimension finie du r\'{e}seau dans la direction $y_{3}$~: le
coefficient de transmission pr\'{e}sente, au niveau des bandes passantes,
des oscillations ap\'{e}riodiques de nature interf\'{e}rentielle. Ces
oscillations, de type Fabry-P\'{e}rot, r\'{e}sultent des interf\'{e}rences
entre les ondes r\'{e}fl\'{e}chies par les deux interfaces dudit milieu
effectif fini. Ces interf\'{e}rences n'ont pas lieu dans les bandes d'arr%
\^{e}t en raison des fortes att\'{e}nuations qui les caract\'{e}risent. Au
contraire, les ondes ne sont pas att\'{e}nu\'{e}es dans les bandes passantes
et peuvent donc se propager \`{a} longue distance dans toute l'\'{e}paisseur
du milieu effectif. Leur forte dispersion explique l'ap\'{e}riodicit\'{e}
des oscillations observ\'{e}es. Si l'on tra\c{c}ait le coefficient de
transmission en fonction de $\mathit{\gamma }^{\prime }D$, les oscillations
se p\'{e}riodiseraient avec une p\'{e}riode inversement proportionnelle \`{a}
l'\'{e}paisseur $h$ du milieu effectif. Les positions des pics d'interf\'{e}%
rence sont en effet donn\'{e}es par $\mathit{\gamma }^{\prime }h=j\pi $,
avec $j=0,1,2,...$.

\section*{ Conclusion}

Le cadre g\'{e}n\'{e}ral dans lequel s'inscrivait ce travail \'{e}tait la
diffusion multiple des ondes \'{e}lastiques antiplanes dans les milieux
solides \'{e}lastiques. L'objectif \'{e}tait l'\'{e}tude de la diffusion
multiple par une multitude de fissures antiplanes p\'{e}riodiquement distribu%
\'{e}es. Au cours de la premi\`{e}re partie de ce travail, l'\'{e}tude de la
diffusion par une fissure antiplane a permis de d\'{e}velopper une \'{e}%
quation int\'{e}grale singuli\`{e}re de type Cauchy. La forme particuli\`{e}%
re de cette \'{e}quation int\'{e}grale se pr\^{e}te \`{a} la mise en \oe %
uvre d'une m\'{e}thode de r\'{e}solution num\'{e}rique adapt\'{e}e \`{a} la d%
\'{e}termination du saut de d\'{e}placement \`{a} travers la fissure. En
employant la formule de repr\'{e}sentation d'une fissure plate et la
condition de p\'{e}riodicit\'{e} d'une rang\'{e}e de fissures situ\'{e}es
sur la m\^{e}me droite antiplane, une \'{e}quation int\'{e}grale de fronti%
\`{e}res a \'{e}t\'{e} obtenue pour le saut de d\'{e}placement \`{a} travers
une fissure dite de r\'{e}f\'{e}rence. Des r\'{e}sultats num\'{e}riques pour
les coefficients de r\'{e}flexion et de transmission ont \'{e}t\'{e} pr\'{e}%
sent\'{e}s en fonction de l'espacement entre fissures, de la fr\'{e}quence
d'excitation et de l'angle d'incidence. La seconde partie de ce travail
avait trait aux r\'{e}seaux bidimensionnels p\'{e}riodiques, constitu\'{e}s
d'un empilement d'un nombre fini de r\'{e}seaux lin\'{e}aires infinis de
fissures en milieu \'{e}lastique. La m\'{e}thode de calcul de la diffusion
par un r\'{e}seau bidimensionnel repose sur la d\'{e}composition du r\'{e}%
seau bidimensionnel en un nombre fini de r\'{e}seaux lin\'{e}aires p\'{e}%
riodiques et infinis qui forment alors des plans r\'{e}ticulaires. La
diffusion par chacun des r\'{e}seaux lin\'{e}aires infinis est d\'{e}termin%
\'{e}e par un calcul exact de diffusion multiple. La diffusion plan apr\`{e}%
s plan est calcul\'{e}e ensuite par une m\'{e}thode it\'{e}rative ou \`{a}
l'aide de la th\'{e}orie de Floquet. Conform\'{e}ment aux th\'{e}ories des
milieux p\'{e}riodiques, la pr\'{e}sence de bandes interdites et permises
dans le spectre des fr\'{e}quences a \'{e}t\'{e} observ\'{e}e. Les effets
dus \`{a} une perturbation de la p\'{e}riodicit\'{e} ont \'{e}galement \'{e}t%
\'{e} analys\'{e}s. Cette perturbation consistait \`{a} introduire un d\'{e}%
faut al\'{e}atoire sur la position de chaque r\'{e}seau lin\'{e}aire.
L'information principale qui en ressort est que les bandes d'arr\^{e}t sont
tr\`{e}s sensibles \`{a} ce type de perturbation, disparaissant assez t\^{o}%
t pour un faible d\'{e}sordre introduit.

La th\'{e}orie d\'{e}velopp\'{e}e ici pourra faire l'objet de d\'{e}%
veloppements ult\'{e}rieurs et \^{e}tre appliqu\'{e}e \`{a} un r\'{e}seau
bidimensionnel avec tout type d'ap\'{e}riodicit\'{e}, telle qu'une variation
des tailles des fissures. En autre, en distribuant de fa\c{c}on al\'{e}%
atoire les r\'{e}seaux lin\'{e}aires, le calcul exact des champs moyens r%
\'{e}fl\'{e}chis et transmis par le milieu al\'{e}atoire (les moyennes \'{e}%
tant calcul\'{e}es en r\'{e}alisant un grand nombre de tirages al\'{e}%
atoires sur les positions des r\'{e}seaux lin\'{e}aires) pourra \^{e}tre
compar\'{e} avec des th\'{e}ories dites de milieu effectif (e.g. \cite{Spr}%
). Ces th\'{e}ories probabilistes fournissent les caract\'{e}ristiques de
l'onde plane, dite coh\'{e}rente, qui se propage dans le milieu homog\`{e}ne 
\'{e}quivalent (vu de l'onde coh\'{e}rente) au milieu al\'{e}atoire.

\eject

\section*{ Liste des figures}

\section*{\textit{\ List of figures}}

\begin{center}
\includegraphics[scale=0.75]{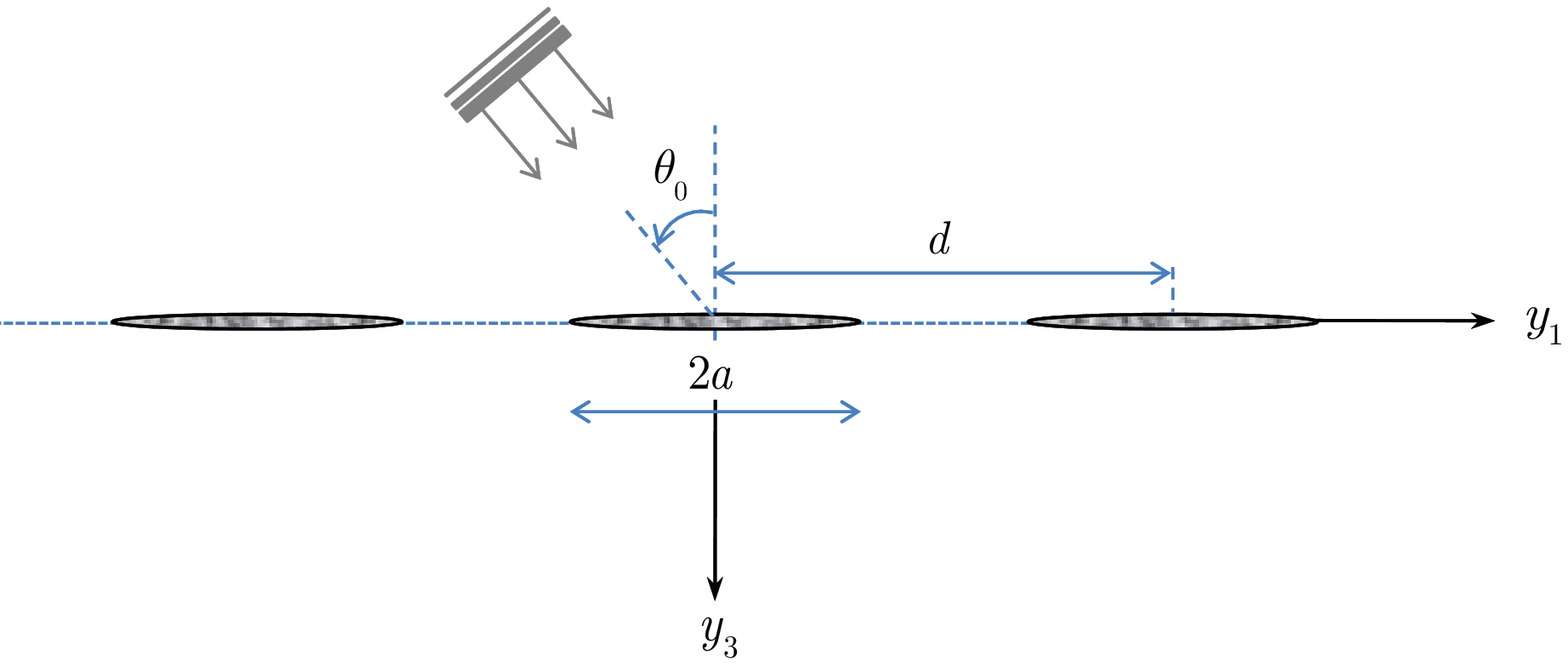}
\end{center}

Fig. 1 : Diffusion d'une onde antiplane par un r\'{e}seau p\'{e}riodique lin%
\'{e}aire de fissures.

\noindent \textit{Fig. 1: Scattering of an antiplane wave by a linear
periodic array of cracks. }

\begin{center}
\includegraphics{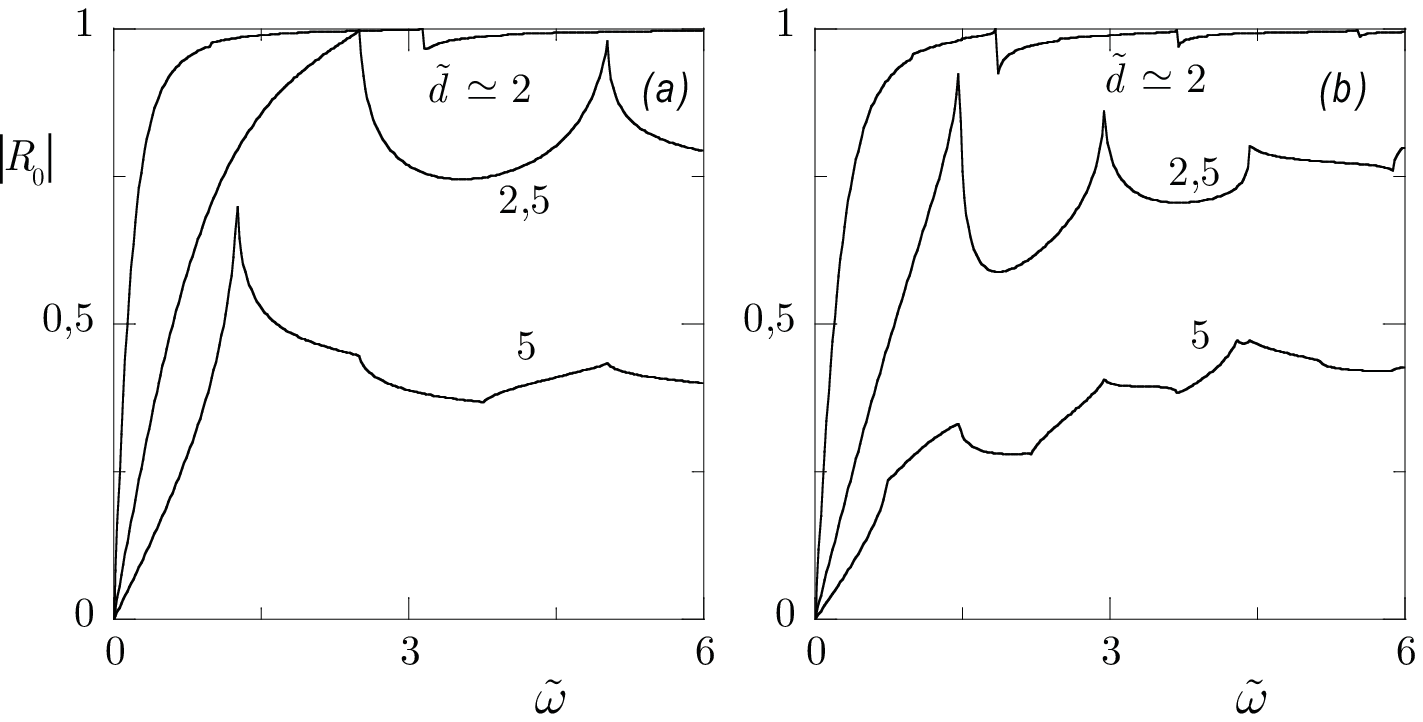}
\end{center}

Fig. 2 : Coefficient de r\'{e}flexion $\left|R_{0} \right|$ versus la fr\'{e}%
quence adimensionnelle $\tilde{\omega }$. Effet de la distance $\tilde{d}$.
a) Incidence normale ($\theta _{0} =0$). b) Incidence oblique ($\theta _{0}
=\pi /4$).

\noindent \textit{Fig. 2: Reflection coefficient $\left|R_{0} \right|$
versus the dimensionless frequency $\tilde{\omega }$. Effect of the distance 
$\tilde{d}$. a) Normal incidence ($\theta _{0} =0$). b) Oblique incidence ($%
\theta _{0} =\pi /4$). }

\begin{center}
\includegraphics{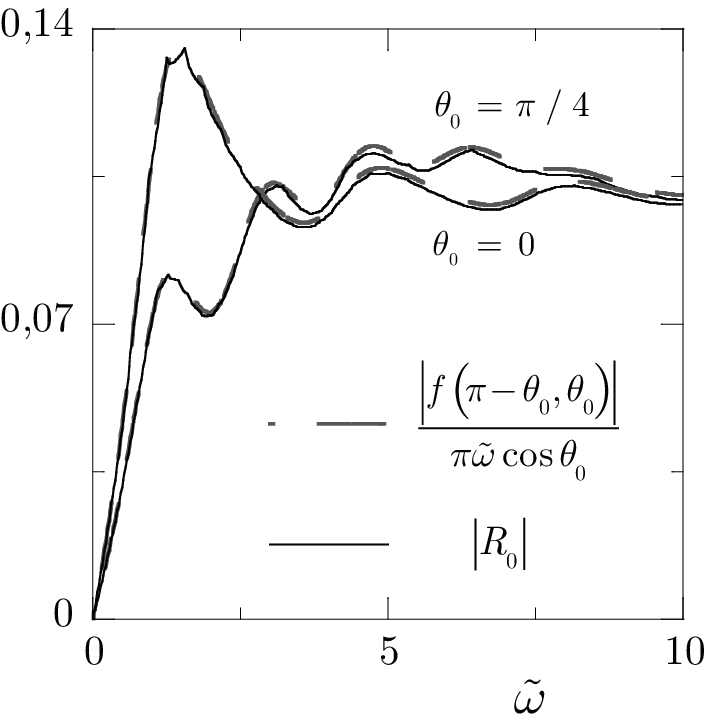}
\end{center}

Fig. 3 : Coefficient de r\'{e}flexion $\left|R_{0} \right|$ versus la fr\'{e}%
quence adimensionnelle $\tilde{\omega }$ pour une distance $\tilde{d}=20$ et
pour deux valeurs de l'angle d'incidence. Comparaison avec le coefficient de
r\'{e}flexion d'une fissure seule.

\noindent \textit{Fig. 3: Reflection coefficient $\left|R_{0} \right|$
versus the dimensionless frequency $\tilde{\omega }$ for a distance $\tilde{d%
}=20$ and for two values of the incidence angle. Comparison with the
reflection coefficient of a single crack.}

\begin{center}
\includegraphics{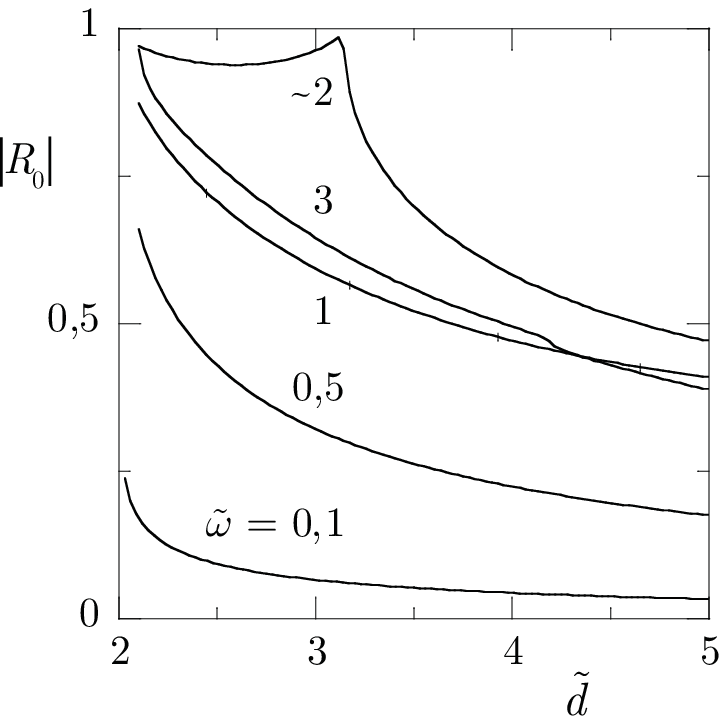}
\end{center}

Fig. 4 : Coefficient de r\'{e}flexion $\left|R_{0} \right|$ versus la
distance $\tilde{d}$ pour $\theta _{0} =0$. Influence de la fr\'{e}quence
adimensionnelle $\tilde{\omega }$.

\noindent \textit{Fig. 4: Reflection coefficient $\left|R_{0} \right|$
versus the distance $\tilde{d}$ for $\theta _{0} =0$. Influence of the
dimensionless frequency $\tilde{\omega }$.}

\begin{center}
\includegraphics[scale=0.75]{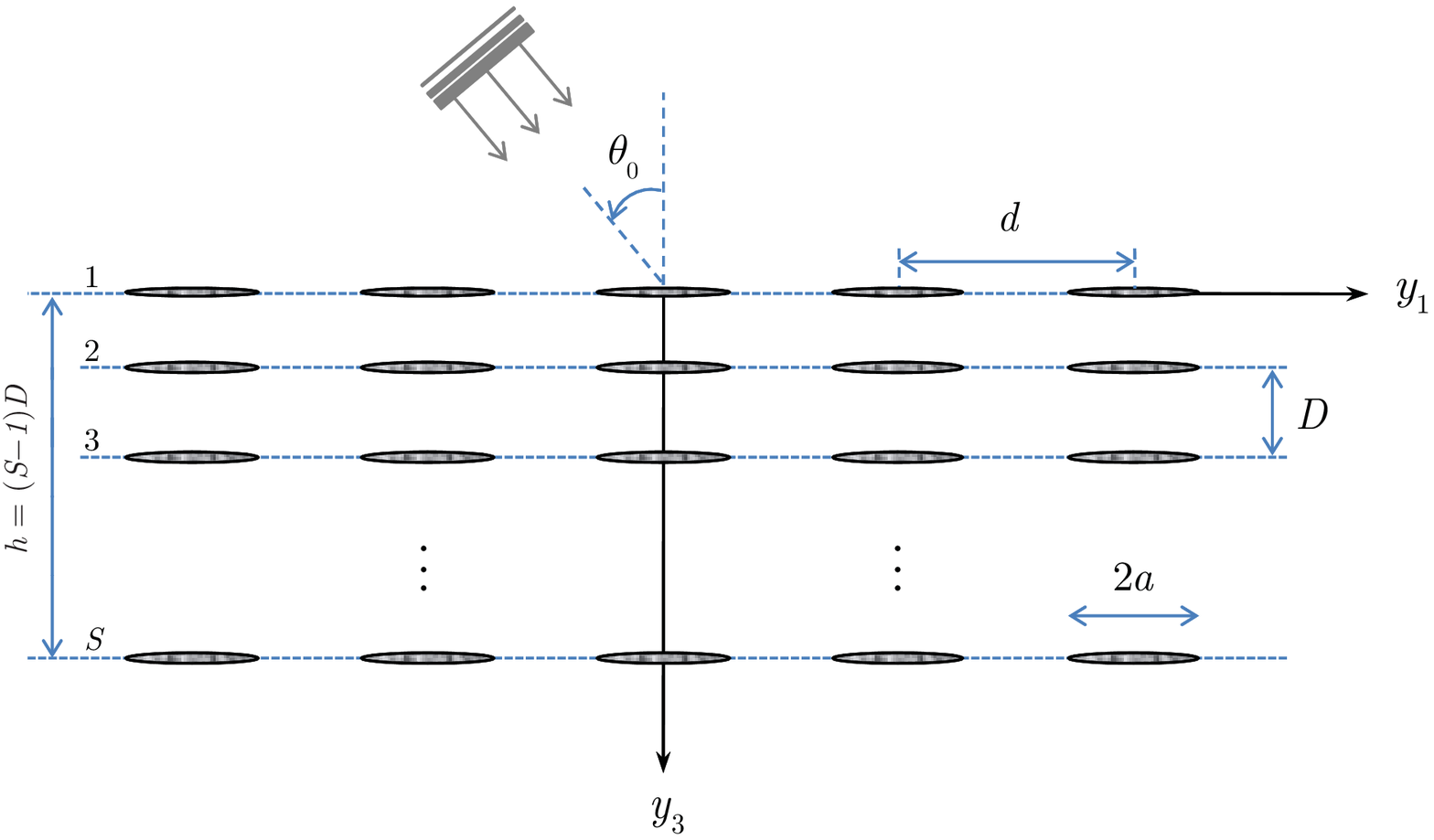}
\end{center}

Fig. 5 : Diffusion d'une onde antiplane par un r\'{e}seau p\'{e}riodique
bidimensionnel de fissures.

\noindent \textit{Fig. 5: Scattering of an antiplane wave by a
two-dimensional periodic array of cracks. }

\begin{center}
\includegraphics{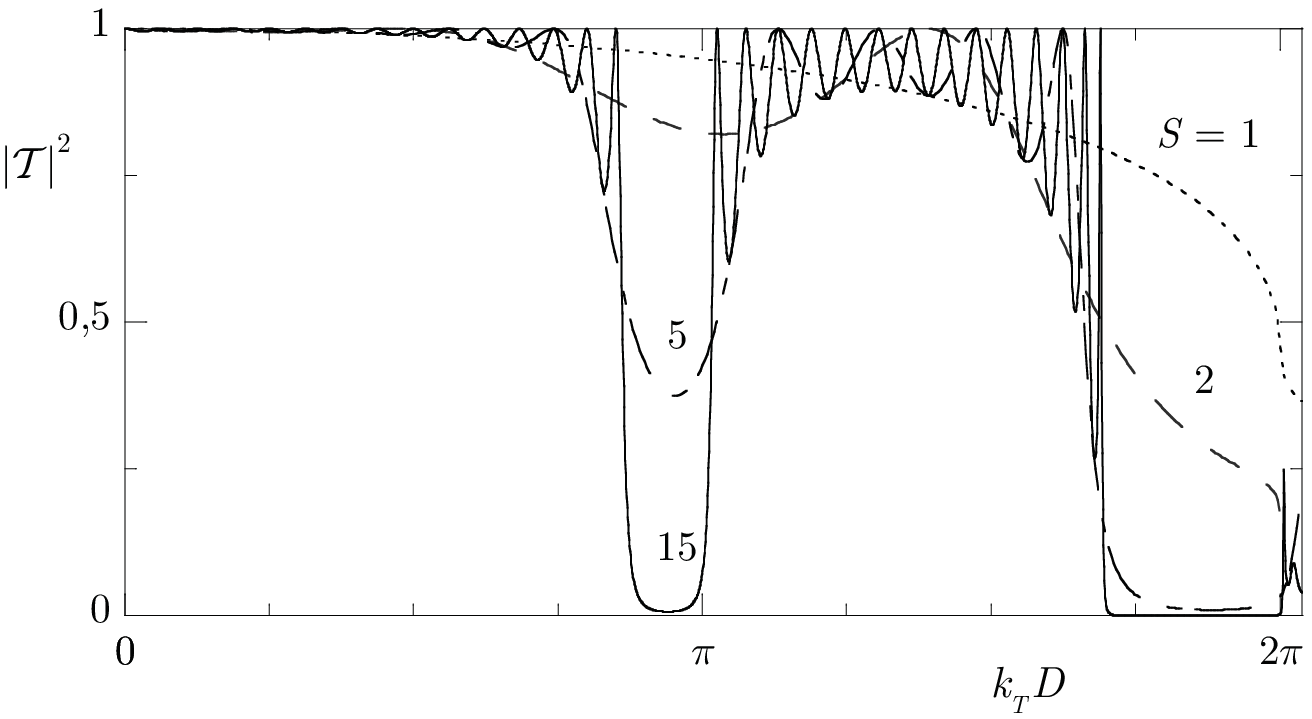}
\end{center}

Fig. 6 : Coefficient de transmission en \'{e}nergie $\left\vert \mathcal{T}%
\right\vert ^{2}$ d'un r\'{e}seau bidimensionnel de fissures excit\'{e} sous
incidence normale avec $\tilde{d}=5$ et $\tilde{D}=1$.

\noindent \textit{Fig. 6: Energy transmission coefficient }$\left\vert 
\mathcal{T}\right\vert ^{2}$\textit{\ of a two-dimensional array of cracks
excited at normal incidence with $\tilde{d}=5$ and $\tilde{D}=1$. }

\begin{center}
\includegraphics{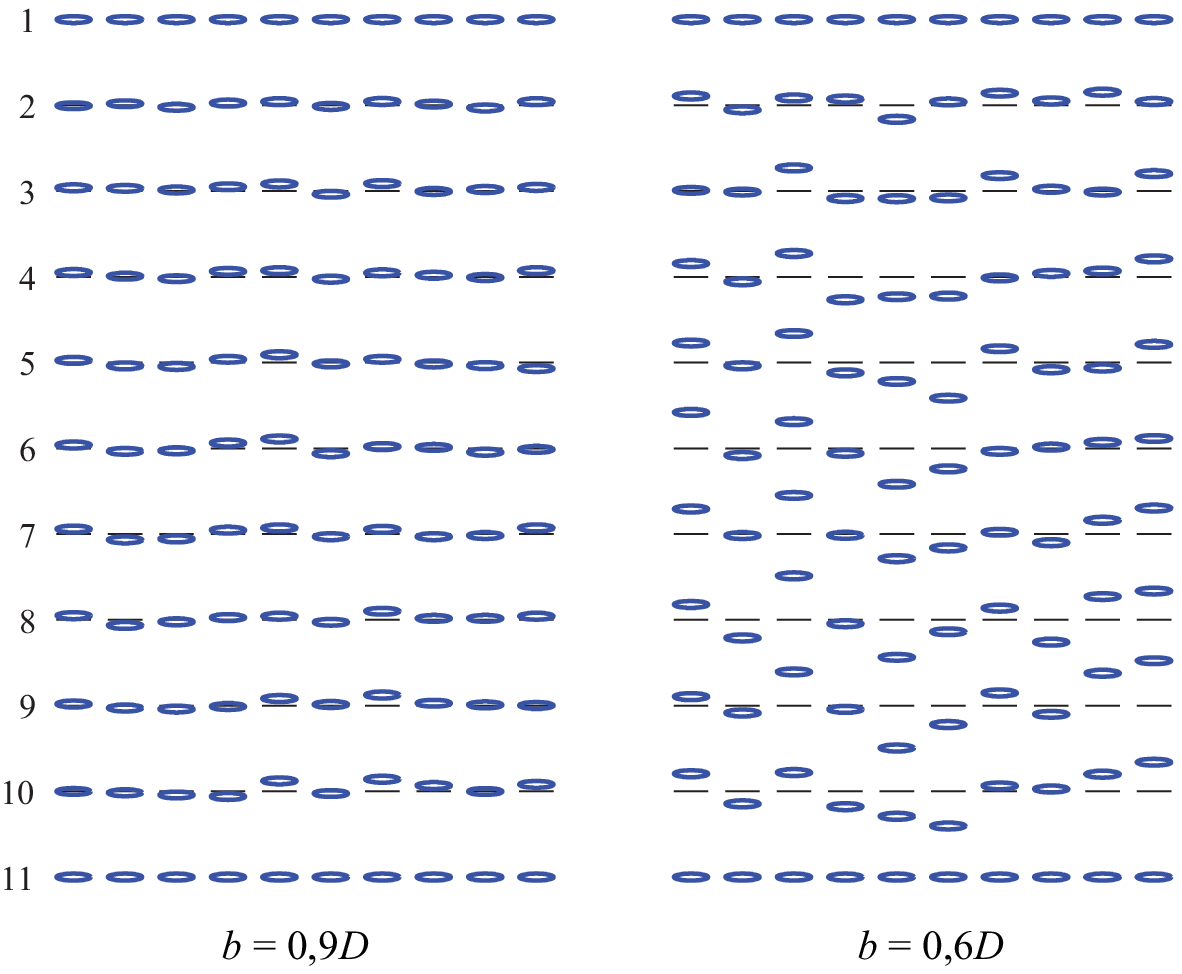}
\end{center}

Fig. 7 : R\'{e}partition des r\'{e}seaux lin\'{e}aires pour deux distances
limites d'approche, $b=0,9D$ et $b=0,6D$, et pour $10$ tirages al\'{e}%
atoires.

\noindent \textit{Fig. 7: }Distribution of linear arrays for two exclusion
distances, $b=0.9D$ and $b=0.6D$, and for $10$ random realisations.

\begin{center}
\includegraphics{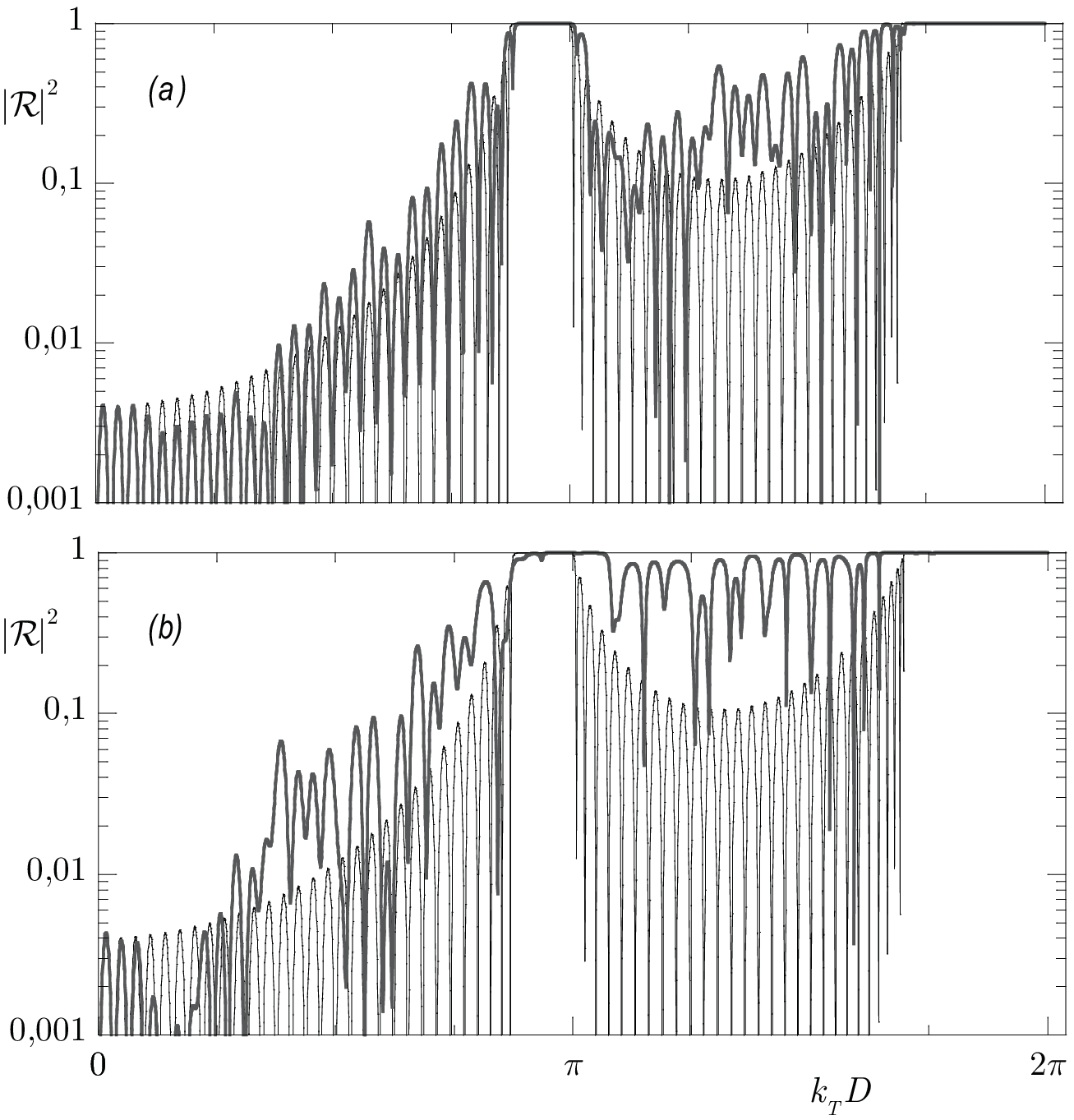}
\end{center}

Fig. 8 : Coefficients de r\'{e}flexion en \'{e}nergie $\left\vert \mathcal{R}%
\right\vert ^{2}$ d'un r\'{e}seau bidimensionnel de fissures avec $\tilde{d}%
=5$ et $\tilde{D}=1$, obtenus pour deux distances limites d'approche : a) $%
b=0,9D$ et b) $b=0,6D$. Ces courbes sont compar\'{e}es avec celle obtenue
pour $b=D$, c'est-\`{a}-dire en situation p\'{e}riodique (trait fin).

\noindent \textit{Fig. 8: Energy reflection coefficients }$\left\vert 
\mathcal{R}\right\vert ^{2}$\textit{\ of a two-dimensional array of cracks
with $\tilde{d}=5$ and $\tilde{D}=1$, obtained for two exclusion distances:
a) }$b=0.9D$ and \textit{b)} $b=0.6D$. \textit{These curves are compared
with the one obtained for }$b=D$\textit{, i.e. in periodic arrangement (fine
line).}

\begin{center}
\includegraphics{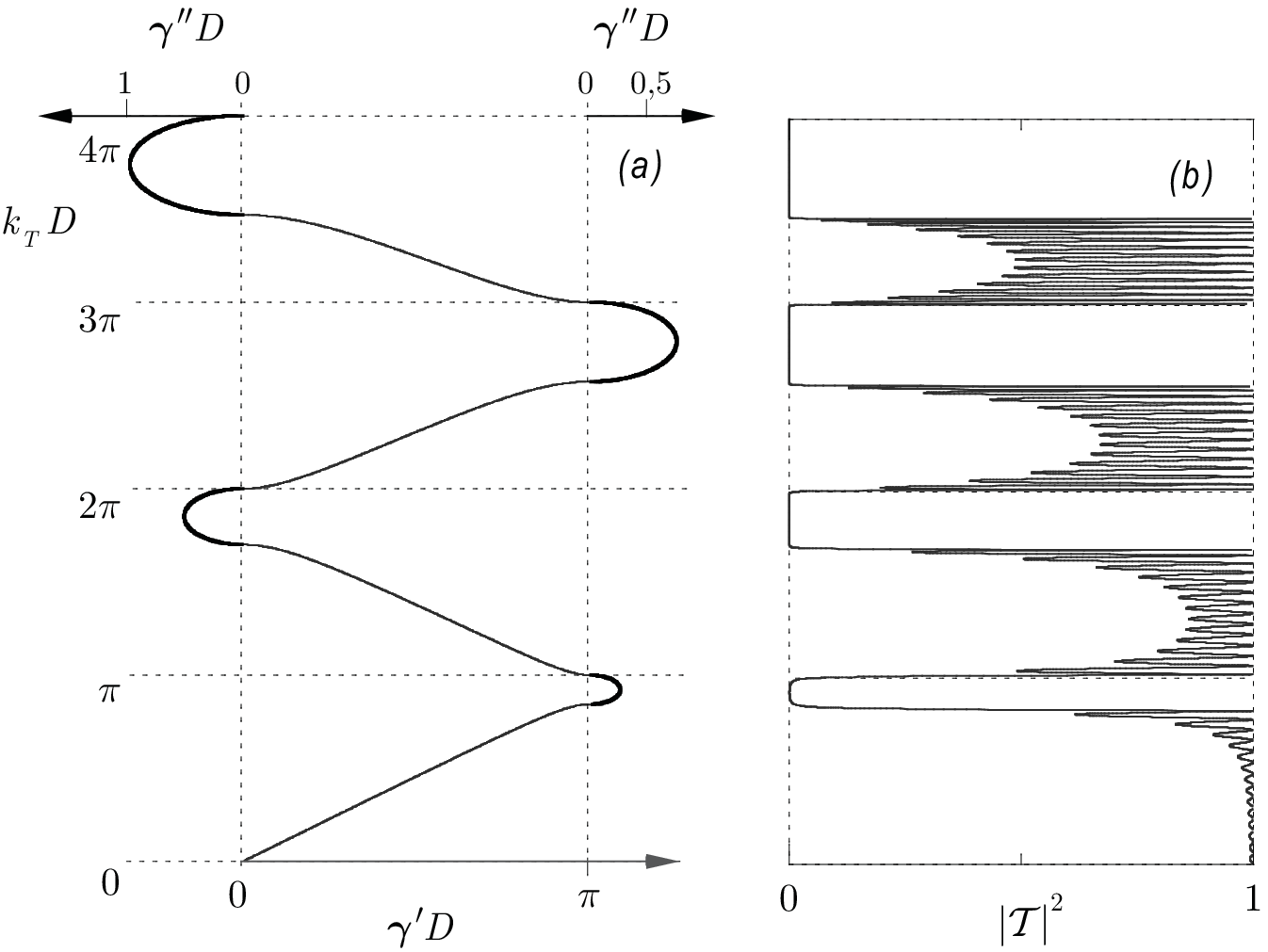}
\end{center}

Fig. 9 : a) Courbes de dispersion (trait fin) et d'att\'{e}nuation (trait
gras) d'un r\'{e}seau bidimensionnel infini de fissures avec $\tilde{d}=2,5$
et $\tilde{D}=4$. b) Coefficient de transmission en \'{e}nergie $\left\vert 
\mathcal{T}\right\vert ^{2}$ d'un r\'{e}seau bidimensionnel fini de fissures.

\noindent \textit{Fig. 9: a) Dispersion (thin line) and attenuation (solid
line) curves of an infinite two-dimensional array of cracks with $\tilde{d}%
=2.5$ and $\tilde{D}=4$. b) Energy transmission coefficient }$\left\vert 
\mathcal{T}\right\vert ^{2}$\textit{\ of a finite two-dimensional array of
cracks.}

\end{document}